\documentclass[12pt,psfig]{article}
\usepackage{amssymb}
\usepackage{amsmath,amsthm}
\usepackage{amsfonts}
\usepackage{graphicx}
\usepackage{braket}
\usepackage{graphics}
\usepackage{indentfirst}
\numberwithin{equation}{section}

\usepackage{hyperref}

\begin{document}
\begin{center}\Large\textbf{Closed String
Radiation from the Interacting
Fractional-Dressed D$p$-Branes with the Transverse Motions}
\end{center}
\vspace{0.75cm}
\begin{center}
{\large Hamidreza Daniali and \large Davoud Kamani}
\end{center}
\begin{center}
\textsl{\small{Department of Physics,
Amirkabir University of
Technology (Tehran Polytechnic), Iran \\
P.O.Box: 15875-4413 \\
e-mails: hrdl@aut.ac.ir , kamani@aut.ac.ir \\}}
\end{center}
\vspace{0.5cm}

\begin{abstract}

We determine the boundary states, associated with a
fractional D$p$-brane
in the presence of the Kalb-Ramond field and a $U(1)$
gauge potential. The background
spacetime is the non-compact orbifold
$ \mathbb{R}^{1, 21} \otimes \mathbb{R}^4/ \mathbb{Z}_2$.
Besides, the brane has a transverse velocity.
We compute the radiation amplitude of a
massless closed string from the interaction of
two fractional D$p$-branes with the previous
background fields.
The branes have been dressed by different fields
and have transverse motions. The total
radiation amplitude will be investigated for
large inter-brane separation.
Finally, the graviton, Kalb-Ramond (axion)
and dilaton emissions will be distinctly studied.
Our calculations are in the context of the
bosonic string theory.

\end{abstract}


{\it PACS numbers}: 11.25.-w; 11.25.Uv

\textsl{Keywords}: Fractional D$p$-branes;
Boundary state; Background fields;
Transverse motion; Radiation amplitude;
Particle emission.

\newpage
\section{Introduction}
\label{100}

The D-branes play a crucial role in the
explanation of string theory \cite{1}, \cite{2}, \cite{3}.
Some of the fundamental techniques in the string theory
have been found by investigating the interactions of
the D-branes. The boundary state formalism is an
adequate techniques for studying the interaction
of the branes in the closed string channel \cite{4}-\cite{14}.
By adding dynamics, various background
and internal fields to the brane,
the most general boundary state, corresponding to
a D$p$-brane has been obtained. The boundary state formalism
has revealed that the dynamical branes
with nonzero background and internal
fields have some interesting properties \cite{15}-\cite{24}.

Among the various configurations of the D-branes,
the fractional D-branes exhibit attractive
behaviors \cite{25}-\cite{37}.
For instance, by using a specific
system of the fractional branes,
the gauge/gravity correspondence
has been investigated \cite{31}-\cite{35}.
Additionally, the fractional branes provide
a deep insight into the Matrix theory \cite{37}-\cite{39}.

In fact, the D-branes are sources of closed strings.
Hence, they can generate closed
strings in a wide range of configurations.
One of them is the production of closed strings
from a solitary unstable D$p$-brane
\cite{40}-\cite{44}.
This kind of production has been investigated in the
presence of various background fields \cite{43, 44}, and its
supersymmetric version has been also studied \cite{45}.
The closed string radiation from the interacting branes
is another important configuration.
This form of radiation has been
studied only in some special configurations
\cite{46}, \cite{47}, \cite{48}.

The background fields, the transverse motions
of the branes and the fractionality of the branes,
motivated and encouraged us to investigate the
effects of these factors on the radiation of
a massless closed string. In our setup, the massless string
is radiated from the interacting parallel D$p$-branes
with background fields
in a partially non-compact orbifoldized spacetime
$\mathbb{R}^{1, 21} \otimes \mathbb{R}^4/ \mathbb{Z}_2$.
Thus, in the context of the bosonic string theory,
we shall use the boundary state formalism to
compute the radiation amplitude.
Hence, by inserting an appropriate
vertex operator into the worldsheet
of the closed string, exchanged between the branes,
we obtain the radiation amplitude.
We shall use the eikonal approximation
in which the recoil of the branes, due to the
string radiation, can be ignored.
Finally, we shall acquire the radiation
amplitude for the case that the distance of the
branes is very large.
We observe that only one of the following
radiations can potentially occur: radiations from each of the
interacting branes and one radiation between the
branes.

This paper is organized as follows.
In Sec. \ref{200}, we shall introduce
the boundary states, corresponding
to the untwisted and twisted sectors of
a fractional-dressed D$p$-brane with a transverse velocity.
In Sec. \ref{301}, the total radiation amplitude for a
massless closed string, resulting from the
interaction of such branes, will be computed. In Sec. \ref{302},
the partition functions and various
correlators in this amplitude
will be explicitly calculated.
In Sec. \ref{303}, the previous amplitude will be
deformed to represent the closed string radiation
from the distant branes.
In Sec. \ref{400}, for each of the graviton,
Kalb-Ramond and dilaton states, we shall explicitly
calculate the total emission amplitude.
In Sec. \ref{500}, the conclusions will be given.

\section{The boundary states}
\label{200}

In this section, we obtain the boundary states
which are corresponding to a
fractional-dressed D$p$-brane. The brane lives in
the $26$-dimensional orbifoldy spacetime
${\mathbb{R}^{1,21}} \times {\mathbb{R}^4}/{\mathbb Z_2}$
and has a transverse motion. The coordinates
of the $\mathbb{R}^4$ are denoted by
$\left\{ x^a|a = 22,23,24,25\right\}$, and
they are influenced by the action of
the $\mathbb{Z}_2$-group. This
group has the cyclic structure $\{e,h | h^2=e\}$.
Under the action of the element $h$ we have the
identification $x^a \equiv -x^a$.
The D$p$-brane has been located
at the fixed points of the orbifold, i.e. $x^a =0$,
which is a $22$-dimensional hyperplane.
Thus, the dimension of a moving brane is restricted by
the upper bound $p\leq 20$.

In order to determine the boundary state,
associated with a fractional
D$p$-brane with background fields, we
consider the following closed string action
\begin{equation}
\label{2.1}
S = -\dfrac{1}{4 \pi \alpha^\prime} \int_\Sigma
d^2 \sigma \left(\sqrt{-h} h^{AB} G_{\mu\nu}
+ \epsilon^{AB} B_{\mu\nu} \right)
\partial_A X^\mu \partial_B X^\nu +
\dfrac{1}{2\pi \alpha^\prime} \int_{\partial\Sigma}
d\sigma A_\alpha \partial_\sigma X^\alpha .
\end{equation}
In this action, we use the following notations:
$\mu$ and $\nu$ are the spacetime
indices and $h^{AB}$, with $A,B\in \{\tau,\sigma\}$,
denotes the metric of the string worldsheet,
while $G_{\mu\nu}$ indicates the spacetime metric
and $B_{\mu\nu}$ represents the
Kalb-Ramond field. Besides, $A_\alpha$
is the gauge potential.
The set $\{X^\alpha| \alpha = 0 ,1, \cdots , p\}$
shows the directions along the
D$p$-brane worldvolume.
We restrict ourselves to the flat spacetime
$G_{\mu\nu} = \eta_{\mu\nu} =
{\rm diag} (-1 , 1,\dots,1)$, flat worldsheet and a
constant Kalb-Ramond field. In addition,
we adopt the usual gauge $A_\alpha = -\frac{1}{2}
F_{\alpha\beta} X^\beta$ with the constant
field strength $F_{\alpha\beta}$.

Apart from the equation of motion,
the following boundary state equations
are also acquired by vanishing the
variation of the action \eqref{2.1},
\begin{eqnarray}
\left(\partial_\tau X^\alpha +\mathcal{F}^{\alpha}_{\ \ \beta}
\partial_\sigma X^\beta \right)_{\tau=0} | B_x\rangle
= 0 ,
\label{2.2}
\end{eqnarray}
\begin{eqnarray}
\left(X^I - y^I\right)_{\tau=0} | B_x\rangle = 0,
\label{2.3}
\end{eqnarray}
where $\mathcal{F}_{\alpha\beta} = F_{\alpha\beta}
- B_{\alpha\beta}$
is the total field strength. The set
$\{x^I| I = p+1 ,\cdots, 25\}$ belongs to the
perpendicular directions of the D$p$-brane worldvolume.
Hence, the parameters $y^I$ indicate the brane's position.

For the non-orbifoldy directions,
the solution of the equation of motion takes the form
\begin{equation}
\label{2.4}
X^\rho (\sigma,\tau) = x^\rho + 2\alpha^\prime p^\rho \tau +
\frac{i}{2} \sqrt{2 \alpha^\prime} \sum_{m\ne 0} \frac{1}{m}
\left(\alpha^\rho_m e^{-2im(\tau-\sigma)}
+ \tilde{\alpha}^\rho_m e^{-2im (\tau+ \sigma)}\right).
\end{equation}
Thus, for the twisted sector (T) we have
$\rho \in \{\alpha,i\}$, where
the set $\{x^i | i = p+1 , \cdots, 21\}$ represents the
non-orbifoldy transverse directions to the brane.
Since the D$p$-brane locates on the fixed points
of the non-compact orbifold $\mathbb{R}^4/\mathbb{Z}_2$,
the orbifoldy coordinates of the closed string
possess the following form
\begin{equation}
\label{2.5}
X^a (\sigma , \tau) = \frac{i}{2}\sqrt{2 \alpha^\prime}
\sum_{r\in \mathbb{Z}
+ \tfrac{1}{2}}  \frac{1}{r}
\left( \alpha^a_r e^{-2ir(\tau-\sigma)}
+ \tilde{\alpha}^a_r e^{-2ir (\tau + \sigma)}\right).
\end{equation}
In the twisted sector, Eq. \eqref{2.3} is
split into the following equations
\begin{eqnarray}
\left(X^i - y^i\right)_{\tau=0} | B_x\rangle^{\rm T}
&=& 0,
\nonumber\\
\left(X^a\right)_{\tau=0} | B_x\rangle^{\rm T}
&=& 0.
\label{2.6}
\end{eqnarray}

Now we impose a transverse velocity to
the brane. Since the brane has stuck at
the orbifold fixed-points, it cannot
move along the
orbifold directions. Therefore, the boost direction
should be a member of the set
$\{x^{p+1}, \cdots, x^{21}\}$,
which we call it $x^{i_v}$.
Hence, Eqs. \eqref{2.2}, \eqref{2.3} and \eqref{2.6},
under the boost effect, take the forms
\begin{eqnarray}
&&\left[\partial_\tau (X^0 - v X^{i_v})
+ \mathcal{F}^0_{\ \ \bar{\alpha} }\partial_{\sigma}
X^{\bar{\alpha}}\right]_{\tau=0}
|B_x\rangle^{\rm U/T}= 0,
\nonumber\\
&&\left[\partial_\tau X^{\bar\alpha}
+ \gamma^2 \mathcal{F}^{\bar\alpha}_{\ \ 0}
\partial_\sigma (X^0 - v X^{i_v})
+ \mathcal{F}^{\bar\alpha}_{\ \ \bar\beta}\partial_\sigma
X^{\bar\beta}\right]_{\tau=0}
|B_x\rangle^{\rm U/T}= 0,
\nonumber\\
&&\left( X^{i_v} - v X^0 - y^{i_v}\right)_{\tau=0}
|B_x\rangle^{\rm U/T}= 0,
\nonumber\\
&& \left( X^{I} -  y^{I}\right)_{\tau=0}
|B_x\rangle^{\rm U}= 0 ,\qquad I \ne i_v,
\nonumber\\
&& \left( X^{i} -  y^{i}\right)_{\tau=0}
|B_x\rangle^{\rm T}= 0 ,\qquad i \ne i_v,
\nonumber \\
&& \left( X^{a} \right)_{\tau=0}
|B_x\rangle^{\rm T}= 0
\label{2.7},
\end{eqnarray}
where $\gamma = 1/\sqrt{1-v^2}$ and
$\{\bar{\alpha}\} = \{\alpha\} - \{0\}$.

By substituting
Eq. \eqref{2.4} and \eqref{2.5} into Eqs. \eqref{2.7},
we get the boundary state equations in terms
of the oscillators and zero modes.
For the oscillating part of the both sectors we obtain
\begin{eqnarray}
&&\left[\alpha^0_m - v \alpha^{i_v}_m -
\mathcal{F}^0_{\ \ \bar\alpha} \alpha^{\bar\alpha}_m
+ \tilde{\alpha}^0_{-m} - v
\tilde{\alpha}^{i_v}_{-m}
+ \mathcal{F}^0_{\ \ \bar\alpha}
\tilde{\alpha}^{\bar\alpha}_{-m}\right]
|B_x\rangle^{\rm U/T}_{\rm osc}
= 0,
\nonumber \\
&&\left\{\alpha^{\bar\alpha}_m - \gamma^2
\mathcal{F}^{\bar\alpha}_{\ \ 0} [\alpha^0_m
- v (\alpha^{i_v}_m -\tilde{\alpha}^{i_v}_{-m})
- \tilde{\alpha}^0_{-m}] +\tilde{\alpha}^{\bar\alpha}_{-m}
-\mathcal{F}^{\bar\alpha}_{\ \ \bar\beta}
(\alpha^{\bar\beta}_{m}
- \tilde{\alpha}^{\bar\beta}_{-m})\right\}
|B_x\rangle^{\rm U/T}_{\rm osc} = 0,
\nonumber\\
&& \left[\alpha^{i_v}_{m} - \tilde{\alpha}^{i_v}_{-m}
-  v  (\alpha^0_m - \tilde{\alpha}^0_{-m})\right]
|B_x\rangle^{\rm U/T}_{\rm osc} = 0,
\nonumber \\
&& \left(\alpha^{i}_{m} - \tilde{\alpha}^i_{-m}\right)
|B_x\rangle^{\rm T}_{\rm osc} = 0 , \qquad i\ne i_v,
\nonumber \\
&& \left(\alpha^{I}_{m} - \tilde{\alpha}^I_{-m}\right)
|B_x\rangle^{\rm U}_{\rm osc} = 0 , \qquad I\ne i_v,
\nonumber \\
&& \left(\alpha^{a}_{m} - \tilde{\alpha}^a_{-m}\right)
|B_x\rangle^{\rm T}_{\rm osc} = 0.
\label{2.8}
\end{eqnarray}
For the zero-mode part, we have
\begin{eqnarray}
&&\left(p^0 - v p^{i_v}\right)
|B_x\rangle^{\rm U/T}_{0} = 0,
\nonumber \\
&&p^{\bar\alpha}|B_x\rangle^{\rm U/T}_{0}= 0,
\nonumber \\
&&\left(x^{i_v} - vx^0 - y^{i_v}\right)
|B_x\rangle^{\rm U/T}_{0} = 0,
\nonumber \\
&&\left(x^i - y^i\right) |B_x\rangle^{\rm T}_{0} = 0 ,
\qquad i\ne i_v,
\nonumber \\
&&\left(x^I - y^I\right) |B_x\rangle^{\rm U}_{0} = 0 ,
\qquad I\ne i_v.
\label{2.9}
\end{eqnarray}

By using the coherent state approach,
the oscillating parts of the boundary states
find the following features
\begin{eqnarray}
|B\rangle^{\rm T}_{\rm osc} &=& \sqrt{- \det M} \exp
\left[-\sum_{m=1}^{\infty} \left(\dfrac{1}{m}
\alpha^\rho_m S_{\rho \rho'}
{\tilde \alpha}^{\rho'}_{-m}
\right)\right] \exp \left[ \sum_{r=1/2}^{\infty}\dfrac{1}{r}
\alpha^a_{-r} \delta_{ab}\tilde{\alpha}_{-r}^b\right]
|0 \rangle
, \ \ \label{2.10}\ \ \\
|B\rangle^{\rm U}_{\rm osc} &=& \sqrt{- \det M}
\exp \left[ - \sum_{m=1}^{\infty} \left(\dfrac{1}{m}
\alpha^\mu_{-m} \hat{S}_{\mu \nu}
{\tilde \alpha}^{\nu}_{-m}\right)\right]|0 \rangle.
\label{2.11}
\end{eqnarray}
The matrices $S$ and ${\hat S}$ are defined by
\begin{eqnarray}
S_{\rho\rho'} &\equiv& \Big(Q_{\lambda\lambda'},
-\delta_{ij} \Big),\;\;i,j \ne i_v ,
\label{2.12}\\
\hat{S}_{\mu\nu} &\equiv& \Big(Q_{\lambda\lambda'}
,-\delta_{ij} , -\delta_{ab} \Big),\;\;i,j \ne i_v ,
\label{2.13}
\end{eqnarray}
where $Q_{\lambda\lambda'}
\equiv(M^{-1}N)_{\lambda\lambda'}$
and $\lambda,\lambda' \in \{\alpha,i_v\}$.
The matrices $M$ and $N$ are given by
\begin{eqnarray}
M^0_{\ \lambda} \equiv \gamma\left(\delta^0_{\ \lambda} -
v \delta^{i_v}_{\ \lambda}-\mathcal{F}^0_{\;\;{\bar\alpha}}
\delta^{\bar\alpha}_{\ \lambda}\right) &\quad,\quad&
N^0_{\ \lambda} \equiv \gamma\left(\delta^0_{\ \lambda}
- v \delta^{i_v}_{\ \lambda}
+ \mathcal{F}^0_{\;\;{\bar\alpha}}
\delta^{\bar\alpha}_{\ \lambda}\right),
\nonumber\\
M^{\bar\alpha}_{\ \lambda} \equiv
\delta^{\bar\alpha}_{\ \lambda}
-\gamma^2 \mathcal{F}^{\bar\alpha}_{\;\;0}
(\delta^{0}_{\ \lambda}
- v\delta^{i_v}_{\ \lambda})
- \mathcal{F}^{\bar\alpha}_{\;\;{\bar\beta}}
\delta^{\bar\beta}_{\ \lambda}&
\quad,\quad & N^{\bar\alpha}_{\ \lambda}
\equiv \delta^{\bar\alpha}_{\ \lambda}
+ \gamma^2 \mathcal{F}^{\bar\alpha}_{\;\;0}
(\delta^{0}_{\ \lambda}
- v\delta^{i_v}_{\ \lambda})
- \mathcal{F}^{\bar\alpha}_{\;\;{\bar\beta}}
\delta^{\bar\beta}_{\ \lambda},
\nonumber\\
M^{i_v}_{\ \lambda} \equiv \delta^{i_v}_{\ \lambda}
- v \delta^{i_v}_{\ \lambda}&\quad,\quad &
N^{i_v}_{\ \lambda}
\equiv -\delta^{i_v}_{\ \lambda} + v
\delta^{i_v}_{\ \lambda}.
\label{2.14}
\end{eqnarray}
The normalization factor in Eqs. \eqref{2.10}
and \eqref{2.11}
are derived from the disk partition
function \cite{4}, \cite{11}.

Applying the quantum mechanical technics,
the zero-mode components of the
boundary states find the following solutions
\begin{eqnarray}
|B_x\rangle_0^{\rm T}
= \delta(x^{i_v} - vx^0 - y^{i_v})
|p^{i_v}=0\rangle \prod_{i=p+1 ,i \ne i_v}^{21}
\Big[\delta(x^i - y^i)
|p^i=0\rangle\Big]\prod_{\alpha=0}^p |p^{\alpha}
=0 \rangle ,
\label{2.15} \\
|B_x\rangle_0^{\rm U}
= \delta(x^{i_v} - vx^0 - y^{i_v})
|p^{i_v}=0\rangle \prod_{I=p+1 ,I \ne i_v}^{25}
\Big[\delta(x^I - y^I)
|p^I=0\rangle\Big]\prod_{\alpha=0}^p |p^{\alpha}
=0\rangle.
\label{2.16}
\end{eqnarray}
For the next purposes, we employ the
integral versions of the Dirac $\delta$-functions, which yield
\begin{equation}
\label{2.17}
|B_x\rangle_0^{\rm T}= \int_{-\infty}^{+\infty}
\prod_{i = p+1}^{21} \dfrac{d p^i}{2\pi}
\exp\left(-ip^i y_i\right)\prod_\rho |p^\rho\rangle,
\end{equation}
\begin{equation}
\label{2.18}
|B_x\rangle_0^{\rm U}= \int_{-\infty}^{+\infty}
\prod_{I = p+1}^{25} \dfrac{d p^I}{2\pi}
\exp\left(-ip^I y_I\right) \prod_\mu |p^\mu\rangle.
\end{equation}
According to Eq. \eqref{2.9} we have
$p^0 = vp^{i_v}$, $p^{\bar \alpha}=0$, and
the remaining components of the momentum in each
sector are nonzero.

The following direct product represents the total
boundary state, associated with the D$p$-brane
\begin{equation}
\label{2.19}
|B\rangle^{\rm U/T}_{\rm tot}= \frac{T_p}{2}
|B_x\rangle^{\rm U/T}_{\rm osc} \otimes |
B_x\rangle^{\rm U/T}_{0} \otimes
|B\rangle_{\rm g},
\end{equation}
where $|B\rangle_{\rm g}$ is the boundary state of the
conformal ghosts
\begin{equation}
\label{2.20}
{|B\rangle}_{\rm g}
=\exp\left[\sum^{\infty}_{m=1}
{\left(c_{-m}{\tilde{b\ }}_{-m}-b_{-m}
{\tilde{c}}_{-m}\right)}\right]\frac{c_0+{\tilde{c}}_0}{2}
\ |q=1\rangle\ \otimes|\tilde{q}=1\rangle.
\end{equation}
Since the ghost fields have no interaction with the
matter fields, their contribution to the boundary states
is not affected by the background
fields, the motion of the brane
and the orbifold projection.

\section{Radiation of a massless closed string}
\label{300}

In this section, we compute the massless closed string
radiation from the interaction of two parallel
D$p$-branes with the background fields and
transverse motions in the spacetime
$\mathbb{R}^{1,21} \otimes \mathbb{R}^4/\mathbb{Z}_2$.
To generalize our computations, let us suppose
that the fields and velocities
of the two D$p$-branes are different.
For showing this difference,
we shall use the subscripts 1 and 2.

\subsection{The total radiation amplitude}
\label{301}

Using the relevant vertex operator,
we can obtain the radiation
of a closed string state. From the mathematical point
of view, one should compute the amplitudes
\begin{equation}
\label{3.1}
\mathcal{A}^{\rm U/T}
= \int_{0}^{\infty} {\rm d}t \int_{0}^t {\rm d}\tau
\int_0^\pi d\sigma \
^{\rm U/T}_{\rm tot}\langle B_1|
e^{-tH^{\rm U/T}} V(\tau, \sigma)
|B_2\rangle_{\rm tot}^{\rm U/T},
\end{equation}
where $V(\tau, \sigma)$ is the vertex operator
of the radiated massless state,
$H^{\rm T}$ and $H^{\rm U}$ are the closed string
Hamiltonians in the twisted and untwisted sectors,
respectively, which are
\begin{eqnarray}
H^{\rm T} &=& H_{\rm g}
+ \alpha^\prime p^\rho p_\rho
+ 2 \sum_{n=1}^\infty
\left(\alpha_{-n}^\rho \alpha_{n\rho}
+ \tilde{\alpha}_{-n}^\rho \tilde{\alpha}_{n\rho}\right)
+ 2 \sum_{r=1/2}^\infty \left( \alpha_{-r}^a \alpha_{ra}
+ \tilde{\alpha}_{-r}^a \tilde{\alpha}_{ra}\right)- 3
\label{3.2},
\qquad \\
H^{\rm U} &=& H_{\rm g} + \alpha^\prime p^\mu p_\mu
+ 2 \sum_{n=1}^\infty
\left( \alpha_{-n}^\mu \alpha_{n\mu}
+ \tilde{\alpha}_{-n}^\mu \tilde{\alpha}_{n\mu}\right)-4 .
\label{3.3}
\end{eqnarray}

Note that, in Eq. \eqref{3.1}
we employed the integrated version of the vertex operator,
which is independent of the conformal ghosts \cite{1, 49}.
It has been demonstrated that for
calculating the bosonic emission 
amplitude, insertion of either
the integrated form of the closed string
vertex operator or its unintegrated
form gives rise to the same result
\cite{1}, \cite{49}. However, this equivalence
will fail for the supersymmetric configurations with certain
compactifications \cite{50}-\cite{53}.
Our setup does not belong to this set of the configurations.
Hence, we have this equivalence, and
the conformal ghosts do not appear in the integrated
vertex operator. Their contributions only appear in
the boundary state ${|B\rangle}_{\rm g}$ and
the Hamiltonian $H_{\rm g}$.

Let us define $z=\sigma +i \tau$ and $\partial = \partial_z$.
The vertex operator in Eq. \eqref{3.1} for a general
massless closed string has the form
\begin{equation}
\label{3.4}
V(z, \bar{z}) = \epsilon_{\mu\nu}
\partial X^\mu \bar{\partial}
X^\nu e^{ip\cdot X},
\end{equation}
where $\epsilon_{\mu\nu}$ represents the polarization tensor,
and ``$p$'' (with $p^2=0$) shows the momentum of the radiated
massless closed string.
Besides, we shall apply the momenta $k_1$ and $k_2$
for the emitted closed string from the
first brane and the absorbed one by
the second brane, respectively.
According to the decomposition
$X^\mu = X^\mu_{0}+ X^\mu_{{\rm osc}}$,
we have 
\begin{eqnarray}
\bar{\partial} X^\mu \partial X^\nu e^{ip\cdot X}
&=& \big[ \bar{\partial}
X^\mu_{0} \partial X^\nu_{0} +\bar{\partial}
X^\mu_{0} \partial X^\nu_{{\rm osc}}
\nonumber \\
&+& \bar{\partial} X^\mu_{{\rm osc}} \partial X^\nu_{0}
+ \bar{\partial} X^\mu_{{\rm osc}} \partial
X^\nu_{{\rm osc}}\big]
e^{ip\cdot X_{0}} e^{ip\cdot X_{{\rm osc}}}.
\label{3.5}
\end{eqnarray}
Hence, the four terms should be sandwiched between the
boundary states. The terms in the vertex
operator may be expressed in a generic form
$A(X_0) e^{i p. X_0} \times B(X_{\rm osc})
e^{i p\cdot X_{\rm osc}}$,
where $A(X_0) \in \{1, \bar{\partial}X_{0} \partial
X_{0}, \partial X_{0}, \bar{\partial}X_{0}\}$
and $ B(X_{\rm osc})\in\{1, \bar{\partial}X_{\rm osc}
\partial X_{\rm osc}, \partial X_{\rm osc},
\bar{\partial}X_{\rm osc}\}$. With this notation we obtain
\begin{eqnarray}
\label{3.6}
^{\rm U/T}_{\rm tot}\langle B_1| e^{-tH^{\rm U/T}}
V(\tau, \sigma)|B_2\rangle_{\rm tot}^{\rm U/T}
&=& \frac{T_p^2}{4} \epsilon_{\mu\nu}
\Big[\ ^{\rm U/T}_{0}\langle B_1 |
e^{-tH^{\rm U/T}(X_{0})}A^{\rm U/T}(X_{0})
e^{ip.X^{\rm U/T}_{0}}|B_2 \rangle_{0}^{\rm U/T}
\nonumber\\
&\times& ^{\rm U/T}_{\rm{osc,g}}\langle
B_1|e^{-t H^{\rm U/T}{\rm(osc,g)} }
B^{\rm U/T} ({X_{\rm osc}})e^{ip\cdot X_{\rm osc}}
|B_2\rangle_{\rm{osc,g}}^{\rm U/T}\Big]^{\mu\nu}.\qquad
\end{eqnarray}

In the twisted sector, first consider
$A^{\rm T}(X_{0}) =1 $,
\begin{eqnarray}
^{\rm T}_{0}\langle B_1 |e^{-tH^{\rm T}(X_{0})}
e^{ip\cdot X^{\rm T}_{0}}|B_2 \rangle_{0}^{\rm T}
&=& \int_{-\infty}^{+\infty}
\int_{-\infty}^{+\infty}e^{3t}\prod_{i
= p+1}^{21}\dfrac{d k_1^i}{2\pi}
\dfrac{d k_2^i}{2\pi}
e^{i k_1^i y_{1 i}} e^{- i k_2^i y_{2 i}}
\nonumber \\
&\times & e^{-t \alpha' k_1^\rho k_{1 \rho}}
e^{2 i \tau \alpha' p_\rho k^\rho_2 } \
\prod_\rho \langle k_1^\rho | (k_2+p)^\rho\rangle .
\label{3.7}
\end{eqnarray}
Using $ \delta(ax) = \frac{1}{|a|}
\delta(x)$, we find
\begin{eqnarray}
\prod_\rho\langle k_1^\rho | (k_2+p)^\rho\rangle
&=& (2\pi)^{22}
\delta (k_2^0+ p^0 - k_1^0) \delta (k_2^{i_v}+ p^{i_v}
- k_1^{i_v}) \prod_{\bar\alpha =1}^{p}
\delta (p^{\bar \alpha}) \prod_{i \ne i_v}^{21}
\delta (k_2^{i}+ p^{i} - k_1^{i})
\nonumber \\
&=& \dfrac{(2\pi)^{22}}{|v_1 - v_2|}
\delta \left[ k_1^{i_v} - \frac{p^0}{v_1 - v_2}
\left( 1 - v_2 \frac{p^{i_v}}{p^0}\right)\right]
\prod_{\bar\alpha =1}^{p}
\delta (p^{\bar \alpha})
\nonumber
\\ &\times & \delta \left[ k_2^{i_v} - \frac{p^0}{v_1
- v_2} \left( 1 - v_1
\frac{p^{i_v}}{p^0}\right)\right] \prod_{i \ne i_v}^{21}
\delta (k_2^{i}+ p^{i} - k_1^{i}).
\label{3.8}
\end{eqnarray}
Therefore, using the Wick's rotation
$\tau \rightarrow -i \tau $, Eq. \eqref{3.7}
takes the feature
\begin{eqnarray}
\label{3.9}
^{\rm T}_{0}\langle B_1 |e^{-tH^{\rm T}(X_{0})}
e^{ip.X^{\rm T}_{0}}|B_2 \rangle_{0}^{\rm T} &=&
\dfrac{e^{3t } }{(2\pi)^{20-2p}}
\frac{1}{|v_1- v_2|}\prod_{\bar\alpha =1}^p \delta
(p^{\bar\alpha})
\nonumber \\
&\times & \int_{-\infty}^{+\infty}
\prod_{i \ne i_v}^{21} dk_1^i e^{i k_1^i b_i}
e^{- t' \alpha'k_1^\rho k_{1 \rho} }
e^{- \tau \alpha'k_2^\rho k_{2 \rho} },
\end{eqnarray}
in which, after the Wick's rotation, we
introduced another proper time $t' = t-\tau$.
The transverse vector $b = y_1 - y_2$
represents the impact parameter.

The remaining components of $A^{\rm T}(X_{0})$
can be simply written as
\begin{eqnarray}
\label{3.10}
&&^{\rm T}_{0}\langle B_1 |e^{-tH^{\rm T}(X_{0})}
\partial X_0^{{\rm T}\rho}e^{ip\cdot X^{\rm T}_{0}}
|B_2 \rangle_{0}^{\rm T} =
\dfrac{e^{3t } }{(2\pi)^{20-2p}}
\frac{1}{|v_1- v_2|} \prod_{\bar\alpha =1}^p \delta
(p^{\bar\alpha})
\nonumber \\
&&\qquad\qquad\qquad\qquad\qquad\qquad\times
\int_{-\infty}^{+\infty}
\prod_{i \ne i_v}^{21} dk_1^i e^{i k_1^i b_i}
e^{- t' \alpha'k_1^\rho k_{1 \rho} }
e^{- \tau \alpha'k_2^\rho k_{2 \rho} }
\left(\alpha' k_1^\rho \right),
\nonumber \\
&&^{\rm T}_{0}\langle B_1 |e^{-tH^{\rm T}(X_{0})}
\bar\partial X_0^{{\rm T}\rho}e^{ip\cdot X^{\rm T}_{0}}
|B_2 \rangle_{0}^{\rm T} =
\dfrac{e^{3t } }{(2\pi)^{20-2p}}\frac{1}{|v_1
- v_2|} \prod_{\bar\alpha =1}^p \delta
(p^{\bar\alpha})
\nonumber \\
&&\qquad\qquad\qquad\qquad\qquad\qquad\times
\int_{-\infty}^{+\infty}
\prod_{i \ne i_v}^{21} dk_1^i e^{i k_1^i b_i}
e^{- t' \alpha'k_1^\rho k_{1 \rho} }
e^{- \tau \alpha'k_2^\rho k_{2 \rho} }
\left(-\alpha' k_1^\rho\right),
\nonumber \\
&&^{\rm T}_{0}\langle B_1 |e^{-tH^{\rm T}(X_{0})} \bar\partial
X_0^{{\rm T}\rho} \partial X_0^{{\rm T}\rho'}
e^{ip\cdot X^{\rm T}_{0}}|B_2 \rangle_{0}^{\rm T} =
\dfrac{e^{3t } }{(2\pi)^{20-2p}}\frac{1}{|v_1
- v_2|} \prod_{\bar\alpha =1}^p \delta
(p^{\bar\alpha})
\nonumber \\
&&\qquad\qquad\qquad\qquad\qquad\qquad\times
\int_{-\infty}^{+\infty}
\prod_{i \ne i_v}^{21} dk_1^i e^{i k_1^i b_i}
e^{- t' \alpha'k_1^\rho k_{1 \rho} }
e^{- \tau \alpha'k_2^\rho k_{2 \rho} }
\left(- \alpha'^2 k_1^\rho k_1^{\rho'}\right).\qquad\quad
\end{eqnarray}
Using the same method, for the zero modes
of the untwisted sector, we acquire the following expressions
\begin{eqnarray}
\label{3.11}
&&^{\rm U}_{0}\langle B_1 |e^{-tH^{\rm U}(X_{0})} \Big\{1 ,
\partial X_0^{\mu} ,\bar\partial X_0^{\mu} ,
\bar\partial X_0^{\mu}\partial X_0^{\nu} \Big\}^{\rm U}
e^{ip.X^{\rm U}_{0}}|B_2 \rangle_{0}^{\rm U} =
\dfrac{e^{4t } }{(2\pi)^{24-2p}}\frac{1}{|v_1- v_2|}
\nonumber \\
&&\quad \times \prod_{\bar\alpha =1}^p \delta
(p^{\bar\alpha})\int_{-\infty}^{+\infty}
\prod_{I \ne i_v}^{25} dk_1^I e^{i k_1^I b_I}
e^{- t' \alpha'k_1^\mu k_{1 \mu} }
e^{- \tau \alpha'k_2^\mu k_{2 \mu} }
\nonumber \\
&& \quad \times\Big\{1 ,\alpha'k_1^\mu , -\alpha'k_1^\mu ,
- \alpha'^2 k_1^\mu k_1^\nu \Big\}.
\end{eqnarray}

Now we consider the oscillating portion
and define the following general correlator
\begin{equation}
\label{3.12}
\langle B ({X_{\rm osc}})
e^{ip\cdot X_{\rm osc}}\rangle^{\rm U/T}
\equiv \ ^{\rm U/T}_{\rm osc}\langle B_1|
e^{-t H^{\rm U/T}(X_{\rm osc}) }
B({X^{\rm U/T}_{\rm osc}})
e^{ip\cdot X^{\rm U/T}_{\rm osc}}
|B_2\rangle^{\rm U/T}_{\rm osc}
\left(\mathcal{Z}^{\rm U/T}_{\rm osc,g}\right)^{-1},
\end{equation}
where
\begin{eqnarray}
\mathcal{Z}^{\rm U/T}_{\rm osc,g}\equiv \
^{\rm U/T}_{\rm osc}\langle B_1|
e^{-t H^{\rm U/T} (X_{\rm osc}) }
|B_2\rangle^{\rm U/T}_{\rm osc}
\otimes \ _{\rm g}\langle B_1|
e^{-t H_{\rm g} } |B_2\rangle_{\rm g},
\nonumber
\end{eqnarray}
is the oscillating and ghost parts of the partition function.
Furthermore, in both sectors, we shall utilize the
following identities
\begin{eqnarray}
\langle \partial X_{\rm osc} e^{ip\cdot X_{\rm osc}}
\rangle^{\rm U/T} &=& i \langle
\partial X\  p\cdot  X \rangle^{\rm U/T}_{\rm osc}
\langle e^{ip\cdot X_{\rm osc}}
\rangle^{\rm U/T},
\label{3.13}\\
\langle \bar{\partial} X_{\rm osc} e^{ip\cdot X_{\rm osc}}
\rangle^{\rm U/T}
&=& i \langle \bar{\partial} X\  p\cdot  X
\rangle^{\rm U/T}_{\rm osc}
\langle e^{ip\cdot X_{\rm osc}}
\rangle^{\rm U/T},
\label{3.14}\\
\langle \partial X_{\rm osc} \bar{\partial} X_{\rm osc}
e^{ip\cdot X_{\rm osc}}\rangle^{\rm U/T}
&=& \left[ \langle \partial X\
\bar{\partial} X \rangle^{\rm U/T}_{\rm osc}
- \langle \partial X\  p . X
\rangle_{\rm osc}^{\rm U/T}
\langle \bar{\partial} X\  p\cdot  X
\rangle^{\rm U/T}_{\rm osc} \right]\nonumber \\
 &\times& \langle e^{ip\cdot X_{\rm osc}} \rangle^{\rm U/T}.
\label{3.15} \ \qquad \
\end{eqnarray}
Note that the related indices of the corresponding
sector should be exerted.

The new proper time enables us to
modify the integrations as
\begin{equation}
\label{3.16}
\int_{0}^{\infty}{\rm d}t \int_{0}^t {\rm d}\tau
=\int_{0}^{\infty}{\rm d}\tau \int_{0}^{\infty}{\rm d}t^\prime.
\end{equation}
The variables $t'$ and $\tau$ represent the proper times of
the radiated closed string from the
first and second brane, respectively.
Thus, $t' = 0$ denotes the radiation from the first brane,
while $\tau =0$ implies the radiation from
the second brane. When the radiation occurs
from the inter-brane region, the condition is $\tau, t' > 0$.

Finally, the general amplitudes \eqref{3.1} are rewritten
by inserting all combinations of Eq. \eqref{3.5}.
For the untwisted sector we acquire
\begin{eqnarray}
\label{3.17}
\mathcal{A}^{\rm U}
&=& \dfrac{T_p^2}{4 (2\pi)^{24-2p}} \frac{1}{|v_1
- v_2|} \prod_{\bar\alpha =1}^p \delta
(p^{\bar\alpha})\int_0^{\infty}{\rm d}t'
\int_0^\infty {\rm d}\tau \int_0^\pi d\sigma \
\int_{-\infty}^{+\infty}
\prod_{I \ne i_v}^{25}{\rm d}k_1^I e^{i k_1^I b_I}
\nonumber \\
&\times& e^{- t' \alpha'k_1^\mu k_{1 \mu} }
e^{- \tau \alpha'k_2^\mu k_{2 \mu} }
e^{4(t'+\tau)} \mathcal{Z}^{\rm U}_{\rm osc,g}
\langle e^{ip\cdot X_{\rm osc}}
\rangle^{\rm U}
\mathcal{M}_{\rm osc}^{\rm U},
\end{eqnarray}
where
\begin{eqnarray}
\label{3.18}
\mathcal{M}_{\rm osc}^{\rm U} &\equiv&
\epsilon_{\mu\nu} \Big[
\langle\partial X^\mu \bar{\partial}
X^\nu \rangle_{\rm osc}^{\rm U}
- p_\gamma p_\eta \langle \partial X^\mu
X^\gamma \rangle_{\rm osc}^{\rm U}
\langle \bar\partial X^\nu
X^\eta \rangle_{\rm osc}^{\rm U}
+ i \alpha' k_1^\nu p_\gamma \langle \partial X^\mu
X^\gamma \rangle_{\rm osc}^{\rm U}
\nonumber \\
&-&i\alpha' k_1^\mu p_\gamma \langle \bar \partial X^\nu
X^\gamma \rangle_{\rm osc}^{\rm U}
- \alpha'^2 k_1^\mu k_1^\nu \Big].
\end{eqnarray}
The additional indices $\gamma$ and $\eta$ belong to the
spacetime directions.

For the twisted sector we find
\begin{eqnarray}
\label{3.19}
\mathcal{A}^{\rm T}&=&
\dfrac{T_p^2}{4 (2\pi)^{20-2p}}\frac{1}{|v_1
- v_2|} \prod_{\bar\alpha =1}^p \delta
(p^{\bar\alpha})\int_0^{\infty}{\rm d}t'
\int_0^\infty {\rm d}\tau\int_0^\pi d\sigma \
\int_{-\infty}^{+\infty}
\prod_{i \ne i_v}^{21}{\rm d}k_1^i e^{i k_1^i b_i}
\nonumber \\
&\times& e^{- t' \alpha'k_1^\rho k_{1 \rho} }
e^{- \tau \alpha'k_2^\rho k_{2 \rho} }
e^{3(t'+\tau)} \mathcal{Z}^{\rm T}_{\rm osc,g}
\langle e^{ip\cdot X_{\rm osc}}
\rangle^{\rm T}
\mathcal{M}_{\rm osc}^{\rm T},
\end{eqnarray}
in which
\begin{eqnarray}
\label{3.20}
\mathcal{M}_{\rm osc}^{\rm T} &\equiv &
\epsilon_{\rho \rho'} \Big[ \langle\partial
X^\rho \bar{\partial} X^{\rho'}
\rangle_{\rm osc}^{\rm T}
- p_\gamma p_\eta \langle \partial X^\rho
X^\gamma \rangle_{\rm osc}^{\rm T}
\langle \bar\partial X^{\rho'}
X^\eta \rangle_{\rm osc}^{\rm T}
+ i \alpha' k_1^{\rho'}
p_\gamma \langle \partial X^\rho
X^\gamma \rangle_{\rm osc}^{\rm T}
\nonumber \\
&-& i\alpha' k_1^\rho p_\gamma \langle \bar\partial X^{\rho'}
X^\gamma \rangle_{\rm osc}^{\rm T}
- \alpha'^2 k_1^\rho k_1^{\rho'} \Big]
\nonumber \\
&+& \epsilon_{ab} \left[ \langle\partial
X^a \bar{\partial} X^b
\rangle_{\rm osc}^{\rm T}
- p_\gamma p_\eta \langle \partial X^a
X^\gamma \rangle_{\rm osc}^{\rm T}\langle
\bar\partial X^{b}
X^\eta \rangle_{\rm osc}^{\rm T}\right].
\end{eqnarray}
After decomposing the indices $\gamma$ and
$\eta$ to the orbifold and non-orbifold directions,
one can easily deduce that the mixed-indices
correlators vanish.

The total amplitude for radiating a massless closed
string is given by
\begin{equation}
\label{3.21}
\mathcal{A}_{\rm tot} =
\mathcal{A}^{\rm T} + \mathcal{A}^{\rm U}.
\end{equation}
The presence of the $\delta$-function in
both sectors obviously
clarifies that the closed string is radiated
perpendicular to the branes.

\subsection{The partition functions and correlators}
\label{302}

Now we compute the partition functions and correlators
that appeared in Eqs. \eqref{3.17}-\eqref{3.20}.
Since we employed the Wick's
rotation to calculate the zero-mode parts,
the oscillating portion should also be
obtained in this frame.

According to the Hamiltonians \eqref{3.2} and
\eqref{3.3}, the partition functions in both sectors become
\begin{eqnarray}
\mathcal{Z}^{\rm U}_{\rm osc,g} &=&
\sqrt{\det(M_1 M_2)} \prod_{n=1}^{\infty}
\left[{\rm det}_{(p+1)\times(p+1)} \left(\mathbf{1}
- \mathcal{Q} \ q^{2n}\right)\right]^{-1}
\left( 1 - q^{2n} \right)^{p-23},
\label{3.22} \\
\mathcal{Z}^{\rm T}_{\rm osc,g} &=&
\sqrt{\det(M_1 M_2)} \prod_{n=1}^{\infty}
\left[{\rm det}_{(p+1)\times(p+1)} \left(\mathbf{1}
- \mathcal{Q} \ q^{2n}\right)\right]^{-1}
\dfrac{\left( 1 - q^{2n} \right)^{p-19}}
{\left( 1 - q^{2n -1} \right)^{4}},
\label{3.23}
\end{eqnarray}
where $q \equiv e^{-2t}$ and $\mathcal{Q} \equiv
Q_1^{\rm T} Q_2$.

For the correlators with a single derivative,
we obtain the following expressions
\begin{eqnarray}
\langle \partial X^\alpha X^\beta
\rangle^{\rm U}_{\rm osc}
&=& i \alpha' \sum_{n=0}^\infty
\Bigg\{{\rm Tr}_{26\times 26}
\left( \dfrac{\eta^{\alpha\beta}
+ Q_1^{{\rm T} \alpha \delta}
Q_{2 \ \delta}^{\beta} (\mathbf{1} -\mathcal{Q}) q^{2(n+1)}}
{\mathbf{1}- \hat{\mathcal{S}} q^{2(n+1)}}\right)
\nonumber \\
&-& Q_2^{\alpha\beta} \mathcal{R}_{(n)}^{{\rm U} \ t'}
+ Q_1^{{\rm T} \alpha\beta}
\mathcal{R}_{(n)}^{{\rm U} \ \tau}
+ Q_2^{\alpha\beta}\mathfrak{R}^{{\rm U} \ t'}_{(n)}
- Q_1^{{\rm T} \alpha\beta}\mathfrak{R}^{U \ \tau}_{(n)}
\Bigg\},
\label{3.24}
\end{eqnarray}
\begin{eqnarray}
\langle \partial X^I X^J \rangle^{\rm U}_{\rm osc}
&=& i \alpha' \delta^{IJ} \sum_{n=0}^\infty\Bigg\{
{\rm Tr}_{26 \times 26}
\left( \dfrac{\mathbf{1} + (\mathbf{1}-\mathcal{Q}) q^{2(n+1)}}
{\mathbf{1}- \hat{\mathcal{S}} q^{2(n+1)}}\right)
\nonumber \\
&-&\mathcal{R}_{(n)}^{U \ t'}
+\mathcal{R}_{(n)}^{U \ \tau}
+ \mathfrak{R}^{U \ t'}_{(n)}
-\mathfrak{R}^{U \ \tau}_{(n)}\Bigg\},
\label{3.25}
\end{eqnarray}
for the untwisted sector, and
\begin{eqnarray}
\langle \partial X^\alpha
X^\beta\rangle^{\rm T}_{\rm osc}
&=& i \alpha' \sum_{n=0}^\infty\Bigg\{
{\rm Tr}_{22\times22}
\left( \dfrac{\eta^{\alpha\beta} + Q_1^{{\rm T}\alpha \delta}
Q_{2 \ \delta}^{\beta} (\mathbf{1}-\mathcal{Q}) q^{2(n+1)}}
{\mathbf{1}- \mathcal{S} q^{2(n+1)}}\right)
\nonumber \\
&-& Q_2^{\alpha\beta} \mathcal{R}_{(n)}^{{\rm T}\ t'}
+ Q_1^{T \alpha\beta}\mathcal{R}_{(n)}^{{\rm T}\ \tau}
+ Q_2^{\alpha\beta} \mathfrak{R}^{{\rm T}\ t'}_{(n)}
- Q_1^{{\rm T} \alpha\beta}
\mathfrak{R}^{{\rm T}\ \tau}_{(n)}\Bigg\},
\label{3.26}
\end{eqnarray}
\begin{eqnarray}
\langle \partial X^i X^j\rangle^{\rm T}_{\rm osc}
&=& i \alpha' \delta^{ij}\sum_{n=0}^\infty
\Bigg\{{\rm Tr}_{22\times22}
\left( \dfrac{\mathbf{1} +  (\mathbf{1}-
\mathcal{Q}) q^{2(n+1)}}
{\mathbf{1}- \mathcal{S} q^{2(n+1)}}\right)
\nonumber \\
&-&\mathcal{R}_{(n)}^{{\rm T}\ t'}
+\mathcal{R}_{(n)}^{{\rm T}\ \tau}
+ \mathfrak{R}^{{\rm T}\ t'}_{(n)}
-\mathfrak{R}^{{\rm T}\ \tau}_{(n)}\Bigg\},
\label{3.27}
\end{eqnarray}
\begin{eqnarray}
\langle \partial X^a X^b\rangle^{\rm T}_{\rm osc}
&=&- i \alpha' \delta^{ab} \sum_{n=0}^\infty
{\rm Tr}_{4\times4} \left( \dfrac{q^{2n}
e^{-4t'}}{\mathbf{1}-q^{2n} e^{-4t'}}
- \dfrac{q^{2n} e^{-4\tau}}{\mathbf{1}
-q^{2n} e^{-4\tau}}\right),
\label{3.28}
\end{eqnarray}
for the twisted sector. Here, we defined
$\mathcal{S}^{\rm T} = \mathcal{S} \equiv S_1^{\rm T} S_2$,
$\mathcal{S}^{\rm U} = \hat{\mathcal{S}}
\equiv \hat S_1^{\rm T}\hat S_2$ and
\begin{eqnarray}
\mathcal{R}_{(n)}^{{\rm U/T}\ (\tau, t')}
&\equiv & {\rm Tr}\left[\dfrac{\mathcal{Q} q^{2n}
e^{-4(\tau,t')}}{\mathbf{1}
- \mathcal{S}^{\rm U/T}
q^{2n} e^{-4(\tau,t')}}\left( \mathbf{1}
+ \dfrac{\mathbf{1}}{\mathbf{1}
- \mathcal{S}^{\rm U/T} q^{2n}
e^{-4(\tau,t')}} \right)\right],
\label{3.29}\\
\mathfrak{R}_{(n)}^{{\rm U/T}\ (\tau, t')}
&\equiv & {\rm Tr}
\left[\dfrac{ q^{2n} e^{-4(\tau,t')}}{\mathbf{1}
- \mathcal{S}^{\rm U/T}
q^{2n} e^{-4(\tau,t')}}\left( \mathbf{1}
+ \dfrac{\mathbf{1}}{\mathbf{1}
- \mathcal{S}^{\rm U/T} q^{2n}
e^{-4(\tau,t')}} \right)\right].
\label{3.30}
\end{eqnarray}
One can easily deduce that
$\langle {\bar \partial} X X\rangle^{\rm T/U}_{\rm osc} =
- \langle \partial X X\rangle^{\rm T/U}_{\rm osc}$.
Since $\partial \bar\partial X = 0 $,
the two-derivative correlator can be computed
by taking the derivative of Eqs. \eqref{3.24}-\eqref{3.28}.

The exponential correlator is obtained
via the Cumulant expansion
in the eikonal approximation;
that is, the recoil of the branes is ignored and the branes
are considered to move in straight trajectories.
Thus, the exponential correlators find the forms
\begin{eqnarray}
\langle e^{ip\cdot X_{\rm osc}} \rangle^{\rm U}
&=& \prod_{n=0}^\infty \Bigg\{ {\rm det}_{(p+1)\times(p+1)}
\left( \mathbf{1}- \mathcal
Q q^{2(n+1)} \right)^{\frac{\alpha'}{2(n+1)}
p_\mu p_\nu ({\hat S_2}{\hat S_1})^{\mu\nu}}
\nonumber \\
&\times & \left[
\dfrac{\Big(1-q^{2(n+1)}\Big)^{p_\mu
p_\nu ({\hat S_2}{\hat S_1})^{\mu\nu}}}
{\Big(1 - q ^{2n}
e^{-4\tau}\Big)^{p_\mu p_\nu \hat S_1^{\mu\nu}}
\Big(1 - q ^{2n} e^{-4t'}\Big)^{p_\mu p_\nu
\hat S_2^{\mu\nu}}}\right]^{\frac{\alpha'}{2(n+1)}(25-p)}
\nonumber \\
&\times & {\rm det}_{(p+1)\times(p+1)}
\Big(\mathbf{1} - \mathcal{Q} q^{2n}
e^{-4\tau}\Big)^{-\frac{\alpha'}{2(n+1)}
p_\mu p_\nu \hat S_1^{\mu\nu}}
\nonumber \\
&\times & {\rm det}_{(p+1)\times(p+1)}
\Big(\mathbf{1} - \mathcal{Q} q^{2n}
e^{-4t'}\Big)^{-\frac{\alpha'}{2(n+1)}
p_\mu p_\nu \hat S_2^{ \mu\nu}}
\nonumber \\
&\times & {\rm det}_{26\times 26}\left[
\exp\left(\frac{\alpha' p_\mu p_\nu
\hat S_1^{\mu\nu}}{2(n+1)} (\mathbf{1}
- \hat{\mathcal{S}} q^{2n} e^{-4\tau})^{-1}\right)\right]
\nonumber \\
&\times & {\rm det}_{26\times 26}\left[
\exp\left(\frac{\alpha' p_\mu p_\nu
\hat S_2^{ \mu\nu}}{2(n+1)} (\mathbf{1}
- \hat{\mathcal{S}}
q^{2n} e^{-4t'})^{-1}\right)\right]\Bigg\},
\label{3.31}
\end{eqnarray}
for the untwisted sector, and
\begin{eqnarray}
\langle e^{ip\cdot X_{\rm osc}} \rangle^{\rm T}
&=& \prod_{n=0}^\infty \Bigg\{{\rm det}_{(p+1)\times(p+1)}
\left( \mathbf{1}- \mathcal
Q q^{2(n+1)} \right)^{\frac{\alpha'}{2(n+1)}
p_\rho p_\sigma ({S_2}{S_1})^{\rho\sigma}}
\nonumber \\
&\times& \left[ \dfrac{\Big(1-q^{2(n+1)}\Big)^{(d-p-5)p_\rho
p_\sigma ({S_2}{S_1})^{\rho\sigma}}}{\Big(1
- q ^{2n} e^{-4\tau}\Big)^{(d-p-5)
p_\rho p_\sigma S_1^{\rho\sigma}- 4p_ap_a} }\right.
\nonumber \\
&\times& \left.\dfrac{1}{\Big(1
- q ^{2n} e^{-4t'}\Big)^{(21-p)p_\rho p_\sigma
S_2^{\rho\sigma}
- 4p_ap_a}}\right]^{\frac{\alpha'}{2(n+1)}}
\nonumber \\
&\times & {\rm det}_{(p+1)\times(p+1)}
\Big(\mathbf{1} - \mathcal{Q} q^{2n}
e^{-4\tau}\Big)^{-\frac{\alpha'}{2(n+1)}
p_\rho p_\sigma S_1^{\rho\sigma}}
\nonumber \\
&\times & {\rm det}_{(p+1)\times(p+1)} \Big(\mathbf{1}
- \mathcal{Q} q^{2n} e^{-4t'}\Big)^{-\frac{\alpha'}{2(n+1)}
p_\rho p_\sigma S_2^{ \rho\sigma}}
\nonumber \\
&\times & {\rm det}_{22\times 22}\left[
\exp\left(\frac{\alpha' p_\rho p_\sigma
S_1^{\rho\sigma}}{2(n+1)} (\mathbf{1}
- \mathcal{S} q^{2n} e^{-4\tau})^{-1}\right)\right]
\nonumber \\
&\times & {\rm det}_{22\times 22}\left[
\exp\left(\frac{\alpha' p_\rho p_\sigma
S_2^{ \rho\sigma}}{2(n+1)} (\mathbf{1}
- \mathcal{S} q^{2n} e^{-4t'})^{-1}
\right)\right]\Bigg\}. \label{3.32}
\end{eqnarray}
for the twisted sector.

Inserting all correlators into the untwisted
and twisted parts of
the amplitude \eqref{3.21}, one receives the
radiation amplitude of the massless
closed string.
Because of the long length of the resultant
amplitude, we do not explicitly write it.

In fact, the integrands of the amplitudes
\eqref{3.17} and \eqref{3.19}
are independent of the worldsheet coordinate
$\sigma$. They merely
depend on the coordinates $\tau$ and $t'$
(see Eqs. \eqref{3.22}-\eqref{3.32}). Thus,
the integration over $\sigma$ is trivial.
One can always multiply the amplitudes by
the factor $\pi$ to restore the effect of the
integration over $\sigma$.

\subsection{Large distance branes}
\label{303}

From now on, we study the closed string radiation
from the branes which are located far from each other.
For two D-branes with large distance,
the interaction effectively occurs via
the massless closed strings. Thus, after a sufficiently
long enough time, $ t\rightarrow\infty $,
which is equivalent to
large inter-brane distance, only the
graviton, Kalb-Ramond and dilaton are effectively
exchanged. We should note that
the limit $t\rightarrow\infty$ must be applied
only to the oscillating portion of the amplitude,
namely the correlators and partition functions.

For simplification, we impose the extra
conditions $\mathcal{S} =\hat{\mathcal{S}} =
\mathbf{1}$ to our formulation, which put constraints
on the parameters of our configuration.
These conditions enable us to compare our
results with the results of Ref. \cite{46}.
When the large distance limit is applied, the
partition functions in both sectors find the form
\begin{equation}
\label{3.33}
\mathcal{Z}^{\rm U}_{\rm osc,g}
|_{t\rightarrow\infty}^{\hat{\mathcal{S}}
= \mathbf{1}} = \mathcal{Z}^{\rm T}_{\rm osc,g}
|_{t\rightarrow\infty}^{\mathcal{S}
= \mathbf{1}} = \sqrt{\det(M_1 M_2)}.
\end{equation}
Besides, the correlators of each
sector are given by
\begin{eqnarray}
\langle \partial X^\alpha
X^\beta\rangle^{\rm U}_{\rm osc}
|_{t\rightarrow\infty}^{\hat{\mathcal{S}}
= \mathbf{1}}  &=&- i \alpha' \left( 13\eta^{\alpha\beta}
- Q_2^{\alpha\beta} \mathcal{T}_{t'}^{\rm U}
- Q_1^{T \alpha\beta}\mathcal{T}_{\tau}^{\rm U}\right)
,\label{3.34}\\
\langle \partial X^I X^J\rangle^{\rm U}_{\rm osc}
|_{t\rightarrow\infty}^{\hat{\mathcal{S}}
= \mathbf{1}}  &=& i \alpha'
\delta^{IJ} \left(13 +  \mathcal{T}_{t'}^{\rm U}
+ \mathcal{T}_{\tau}^{\rm U}\right),
\label{3.35}\\
\langle \partial X^\alpha X^\beta
\rangle^{\rm T}_{\rm osc}
|_{t\rightarrow\infty}^{\mathcal{S}
= \mathbf{1}}  &=&- i \alpha'
\left( 11\eta^{\alpha\beta}
- Q_2^{\alpha\beta} \mathcal{T}_{t'}^{\rm T}
- Q_1^{T \alpha\beta} \mathcal{T}_{\tau}^{\rm T}\right)
,\label{3.36}\\
\langle \partial X^i X^j\rangle^{\rm T}_{\rm osc}
|_{t\rightarrow\infty}^{\mathcal{S}
= \mathbf{1}}  &=& i \alpha' \delta^{ij}
\left( 11+ \mathcal{T}_{t'}^{\rm T}
+ \mathcal{T}_{\tau}^{\rm T}\right),
\label{3.37}\\
\langle \partial X^a X^b\rangle^{\rm T}_{\rm osc}
|_{t\rightarrow\infty}^{\mathcal{S}
= \mathbf{1}}  &=&
\langle \partial X^a X^b\rangle^{\rm T}_{\rm osc}
|_{t\rightarrow\infty}  = -4i \alpha' \delta^{ab} \left[
f(\tau) - f(t')\right],\label{3.38}
\end{eqnarray}
where we used these definitions
\begin{eqnarray}
\mathcal{T}_{\tau}^{\rm T/U} &=&
{\rm Tr}^{\rm T/U}
\left[ \frac{e^{-4\tau}}{\mathbf{1}
- e^{-4\tau}} \left( \mathbf{1}
+ \frac{\mathbf {1}}{\mathbf{1}
- e^{-4\tau}}\right)\right],
\label{3.39}\\
\mathcal{T}_{t'}^{\rm T/U}
&=& -{\rm Tr}^{\rm T/U}
\left[ \frac{e^{-4t'}}{\mathbf{1}
- e^{-4t'}} \left( \mathbf{1}
+ \frac{\mathbf {1}}{\mathbf{1}
- e^{-4t'}}\right)\right],
\label{3.40}\\
f(t',\tau) &=& \frac{e^{-4 (t',\tau)}}{1
- e^{(t',\tau)}}.
\label{3.41}
\end{eqnarray}
In the right-hand side we have
${\rm Tr}^{\rm U} \rightarrow 26 $,
 ${\rm Tr}^{\rm T} \rightarrow 22$
and $\mathbf{1} \rightarrow 1 $.

In this limit, the exponential correlators are deformed as
\begin{eqnarray}
\langle e^{ip\cdot X_{\rm osc}}
\rangle^{\rm U}|_{t\rightarrow\infty}^{\hat{\mathcal{S}}
= \mathbf{1}}
&= &  \left[\Big(1 -e^{-4\tau}\Big)^{p_\mu p_\nu
\hat S_1^{\mu\nu}}\Big(1 -  e^{-4t'}\Big)^{p_\mu p_\nu
\hat S_2^{\mu\nu}}\right]^{-13 \alpha'}
\nonumber \\
&\times & \exp \Bigg\{13 \alpha' p_\mu p_\nu
\Bigg(\frac{\hat S_1^{\mu\nu}}{1-e^{-4\tau}}
+ \frac{\hat S_2^{ \mu\nu}}{1-e^{-4t'}}
\nonumber \\
&+& (\hat S_1 + \hat S_2)^{\mu\nu}
(\gamma_{\rm EM}-1)\Bigg)\Bigg\}
\label{3.42}
\end{eqnarray}
for the untwisted sector, and
\begin{eqnarray}
\langle e^{ip\cdot X_{\rm osc}} \rangle_{\rm osc}^{\rm T}
|_{t\rightarrow\infty}^{\mathcal{S}
= \mathbf{1}}
&=&  \left[\Big(1 -e^{-4\tau}\Big)^{22p_\rho p_\sigma
S_1^{\rho\sigma}- 4p_a p_a}
\Big(1 -  e^{-4t'}\Big)^{22p_\rho p_\sigma
S_2^{\rho\sigma} - 4p_a p_a}\right]^{-\alpha'/2}
\nonumber \\
&\times & \exp \Bigg\{11 \alpha' p_\rho p_\sigma
\Bigg(\frac{ S_1^{\rho\sigma}}{1-e^{-4\tau}}
+ \frac{ S_2^{ \rho\sigma}}{1-e^{-4t'}}
\nonumber \\
&+& (S_1 + S_2)^{\rho\sigma} (\gamma_{\rm EM}-1)\Bigg)\Bigg\}
\label{3.43}
\end{eqnarray}
for the twisted sector. The Euler-Mascheroni number
$\gamma_{\rm EM} = 0.577 \dots$
was entered through a regularization scheme.

We can rewrite the two-derivative term as a
combination of one-derivative terms.
The contribution of the two-derivative term
to the untwisted part of the amplitude is given by
\begin{eqnarray}
\label{3.44}
\mathcal{A}^{\rm U}_{\partial \bar\partial}
|_{t\rightarrow\infty}^{\hat{\mathcal{S}}
= \mathbf{1} } &=& \dfrac{T_p^2}{8 (2\pi)^{23-2p}}
\frac{1}{|v_1- v_2|} \prod_{\bar\alpha =1}^p \delta
(p^{\bar\alpha})  \int_0^{\infty} dt' \int_0^\infty d\tau
\int_{-\infty}^{+\infty}
\prod_{I \ne i_v}^{25} dk_1^I
e^{i k_1^I b_I}\epsilon_{\mu\nu}
\nonumber \\
&\times& e^{- t' \alpha'k_1^\mu k_{1 \mu}}
e^{- \tau \alpha'k_2^\mu k_{2 \mu}}
e^{4(t'+\tau)}\mathcal{Z}^{\rm U}_{\rm osc,g}
|_{t\rightarrow\infty}^{\hat{\mathcal{S}} = \mathbf{1}}
\langle e^{ip\cdot X_{\rm osc}}
\rangle^{\rm U}|_{t\rightarrow\infty}^{\hat{\mathcal{S}}
= \mathbf{1}}
\langle\partial X^\mu \bar{\partial}
X^\nu \rangle_{\rm osc}^{\rm U}.
\end{eqnarray}
With the help of integration by part, one finds
\begin{eqnarray}
\mathcal{A}^{\rm U}_{\partial \bar\partial}
|_{t\rightarrow\infty}^{\hat{\mathcal{S}}
= \mathbf{1} } &=&
- \dfrac{T_p^2}{8 (2\pi)^{23-2p}}
\frac{1}{|v_1- v_2|} \prod_{\bar\alpha =1}^p \delta
(p^{\bar\alpha})  \int_0^{\infty} dt' \int_0^\infty d\tau
\int_{-\infty}^{+\infty}
\prod_{I \ne i_v}^{25}
dk_1^I e^{i k_1^I b_I}\epsilon_{\mu\nu}
\nonumber \\
&\times& e^{- t' \alpha'k_1^\mu k_{1 \mu} }
e^{- \tau \alpha'k_2^\mu k_{2 \mu} }
e^{4(t'+\tau)}\mathcal{Z}^{\rm U}_{\rm osc,g}
|_{t\rightarrow\infty}^{\hat{\mathcal{S}} = \mathbf{1} }
\langle e^{ip\cdot X_{\rm osc}}\rangle^{\rm U}
|_{t\rightarrow\infty}^{\hat{\mathcal{S}}
= \mathbf{1} }
\nonumber \\
&\times& \langle\partial X^\mu
X^\nu \rangle_{\rm osc}^{\rm U}
|_{t\rightarrow\infty}^{\hat{\mathcal{S}}
= \mathbf{1} }\Big( p_\gamma p_\eta \langle\partial X^\gamma
X^\eta \rangle_{\rm osc}^{\rm U}
|_{t\rightarrow\infty}^{\hat{\mathcal{S}}
= \mathbf{1} } + \frac{i}{2} \alpha'
(k_1^\mu k_{1 \mu} - k_2^\mu k_{2 \mu})\Big)
\nonumber\\
&+&{\rm Non-intergral\ term}.
\label{3.45}
\end{eqnarray}
From the physical point of view, we should exclude the
non-integral terms. Consequently,
omission of the surface term
at $\tau , t'= 0$ implies that the quantities
$p_\mu p_\nu {\hat S}_1^{\mu\nu}$ and
$p_\mu p_\nu {\hat S}_2^{ \mu\nu}$
should be negative.

For the twisted sector, the two-derivative term
possesses the following contribution to
the amplitude
\begin{eqnarray}
\mathcal{A}^{\rm T}_{\partial \bar\partial}
|_{t\rightarrow\infty}^{\mathcal{S}
= \mathbf{1} } &=&
- \dfrac{T_p^2}{8 (2\pi)^{19-2p}}
\frac{1}{|v_1- v_2|} \prod_{\bar\alpha =1}^p \delta
(p^{\bar\alpha})  \int_0^{\infty} dt' \int_0^\infty d\tau
\int_{-\infty}^{+\infty}  \prod_{i \ne i_v}^{21} dk_1^i
e^{i k_1^i b_i}
\nonumber \\
&\times& e^{- t' \alpha'k_1^\rho k_{1 \rho} }
e^{- \tau \alpha'k_2^\rho k_{2 \rho} }
e^{3(t'+\tau)}\mathcal{Z}^{\rm T}_{\rm osc,g}
|_{t\rightarrow\infty}^{\mathcal{S} = \mathbf{1} }
\langle e^{ip\cdot X_{\rm osc}}
\rangle^{\rm T}|_{t\rightarrow\infty}^{\mathcal{S}= \mathbf{1}}
\nonumber \\
&\times& \Bigg\{ \epsilon_{\rho\sigma} \langle\partial X^\rho
X^\sigma \rangle_{\rm osc}^{\rm T}
|_{t\rightarrow\infty}^{\mathcal{S}
= \mathbf{1} }\Big[ p_{\rho'}
p_{\sigma'} \langle\partial X^{\rho'}
X^{\sigma'} \rangle_{\rm osc}^{\rm T}
|_{t\rightarrow\infty}^{\mathcal{S}
= \mathbf{1} } + \frac{i}{2} \alpha'
(k_1^{\rho'} k_{1 {\rho'}}
- k_2^{\rho'} k_{2 {\rho'}})\Big]
\nonumber \\
&+& \epsilon_{ab} \langle\partial X^a
X^b \rangle_{\rm osc}|_{t\rightarrow\infty}^{\mathcal{S}
= \mathbf{1} }\Big[p_{a'} p_{b'} \langle\partial X^{a'}
X^{b'} \rangle_{\rm osc}|_{t\rightarrow\infty}^{\mathcal{S}
= \mathbf{1} } + \frac{i}{2}
\alpha' (k_1^\rho k_{1 \rho} -
k_2^\rho k_{2 \rho})\Big]\Bigg\}, \qquad
 \label{3.46}
\end{eqnarray}
where the additional constraints are
\begin{eqnarray}
&~& p_\rho p_\sigma S_1^{\rho\sigma} <0 \;\;\;,
\;\;\;p_\rho p_\sigma S_2^{\rho\sigma} < 0 ,
\nonumber\\
&~& (21-p) p_\rho p_\sigma S_1^{\rho\sigma} >4 p_a p_a \;\;\;,
\;\;\;(21-p) p_\rho p_\sigma S_2^{ \rho\sigma} > 4 p_a p_a.
\label{3.47}
\end{eqnarray}

\section{The particles emission}
\label{400}

In this section, we shall calculate the graviton, Kalb-Ramond
and dilaton emissions from the interacting
branes with the large distance.
Instead of writing three different amplitudes
for the three massless states, 
at first for all massless states
we write a general emission amplitude,
which includes several variables.
Then, we explicitly write these variables for each of the
graviton, Kalb-Ramond and dilaton to separate the
previous general amplitude to three distinct amplitudes.
For each sector, the amplitude is given by
\begin{eqnarray}
\mathcal{A}^{\rm U}
|_{t\rightarrow\infty}^{\hat{\mathcal{S}}
= \mathbf{1} }&=& \dfrac{T_p^2\sqrt{\det(M_1 M_2)}}{8
(2\pi)^{23-2p}} \frac{1}{|v_1- v_2|}
\prod_{\bar\alpha =1}^p \delta
(p^{\bar\alpha})  \int_0^{\infty} dt' \int_0^\infty d\tau
\int_{-\infty}^{+\infty}
\prod_{I \ne i_v}^{25} dk_1^I e^{i k_1^I b_I}
\nonumber \\
&\times& e^{- t' \alpha'k_1^\mu k_{1 \mu} }
e^{- \tau \alpha'k_2^\mu k_{2 \mu} }
e^{4(t'+\tau)}\langle e^{ip\cdot X_{\rm osc}}
\rangle^{\rm U}|_{t\rightarrow\infty}^{\hat{\mathcal{S}}
= \mathbf{1} }
\nonumber \\
&\times& \alpha'^2 \left[ \mathbf{A}^{\rm U}
\left(\mathcal{T}_\tau^{\rm U}\right)^2
+ \mathbf{B}^{\rm U}
\left(\mathcal{T}_{t'}^{\rm U}\right)^2
+ \mathbf{C}^{\rm U}
\mathcal{T}_\tau^{\rm U}
+ \mathbf{D}^{\rm U} \mathcal{T}_{t'}^{\rm U}
+ \mathbf E^{\rm U} \right],
\label{4.1}
\end{eqnarray}
for the untwisted sector, and
\begin{eqnarray}
\mathcal{A}^{\rm T}|_{t\rightarrow\infty}^{\mathcal{S}
= \mathbf{1}}&=&\dfrac{T_p^2\sqrt{\det(M_1 M_2)}}
{8 (2\pi)^{19-2p}} \frac{1}{|v_1- v_2|}
\prod_{\bar\alpha =1}^p \delta
(p^{\bar\alpha})  \int_0^{\infty} dt' \int_0^\infty d\tau
\int_{-\infty}^{+\infty}
\prod_{i \ne i_v}^{21} dk_1^i e^{i k_1^i b_i}
\nonumber \\
&\times& e^{- t' \alpha'k_1^\rho k_{1 \rho'} }
e^{- \tau \alpha'k_2^\rho k_{2 \rho'} }
e^{3(t'+\tau)}\langle e^{ip\cdot X_{\rm osc}}
\rangle^{\rm T}|_{t\rightarrow\infty}^{\mathcal{S}
= \mathbf{1}}
\nonumber \\
&\times& \alpha'^2 \Big\{ \mathbf{A}^{\rm T}
\left(\mathcal{T}_\tau^{\rm T}\right)^2
+ \mathbf{B}^{\rm T}\left(\mathcal{T}_{t'}^{\rm T}\right)^2
+ \mathbf{C}^{\rm T}\mathcal{T}_\tau^{\rm T}
+ \mathbf{D}^{\rm T} \mathcal{T}_{t'}^{\rm T}
+ \mathbf E^{\rm T}
\nonumber \\
&+& {\mathbf F} \big[f^2(\tau) + f^2({t'}) \big]
- \mathbf G \big[ f(\tau) - f(t')\big]  \Big\},
\label{4.2}
\end{eqnarray}
for the twisted sector.
Using the integration by part on the proper times,
the following equivalence relations are obtained
\begin{eqnarray}
\mathcal{T}_\tau^{\rm T} \doteq
- \frac{\alpha'k_2^\rho k_{2\rho}
- 3}{2\alpha'p_\rho p_\sigma S_1^{\rho \sigma}}
& ,\qquad & \mathcal{T}_{t'}^{\rm T} \doteq
\frac{\alpha'k_1^\rho k_{1\rho} - 3}{2\alpha'p_\rho
p_\sigma S_2^{ \rho \sigma}},
\label{4.3}\\
\mathcal{T}_\tau^{\rm U} \doteq
- \frac{\alpha'k_2^\mu k_{2\mu}
- 4}{2\alpha'p_\mu p_\nu S_1^{\mu \nu}} & ,\qquad &
\mathcal{T}_{t'}^{\rm U}
\doteq \frac{\alpha'k_1^\mu k_{1\mu}
- 4}{2\alpha'p_\mu p_\nu S_2^{ \mu \nu}},
\label{4.4}\\
f(\tau) \doteq -\frac{\alpha' k^\rho_1 k_{1\rho}- 3}
{2\alpha' p_a p_a}& ,\qquad &
f(t') \doteq -\frac{\alpha' k^\rho_2 k_{2\rho}- 3}{2\alpha'
p_a p_a}.
\label{4.5}
\end{eqnarray}
For example, for the untwisted sector see
the Refs. \cite{46, 47}.
In the rest of this paper we shall explicitly calculate and concentrate on the
variables $\mathbf{A}^{\rm U/T}$,
$\mathbf{B}^{\rm U/T}$, $\mathbf{C}^{\rm U/T}$,
$\mathbf{D}^{\rm U/T}$, $\mathbf E^{\rm U/T}$,
$\mathbf F$ and $\mathbf G$ for the
graviton, Kalb-Ramond and dilaton states.

Now we compute the 
integrals over the proper times in the eikonal
approximation. For the twisted sector, the result is
\begin{eqnarray}
\int_0^{\infty} dt' \int_0^\infty
d\tau e^{- t' \alpha'k_1^\rho k_{1 \rho}}
e^{- \tau \alpha'k_2^\rho k_{2 \rho}}
e^{3(t'+\tau)}\langle e^{ip\cdot X_{\rm osc}}
\rangle^{\rm T}|_{t\rightarrow\infty}^{\mathcal{S}
= \mathbf{1}} = \mathcal{I}_1 \mathcal{I}_2 ,
\label{4.6}
\end{eqnarray}
\begin{eqnarray}
\mathcal{I}_{1,2} &=& \dfrac{1}{4}
\exp \left[11 \alpha' p_\rho p_\sigma
S_{(1,2)}^{\rho\sigma}(\gamma_{\rm EM}-1)\right]
\nonumber \\
&\times & \left\{ \dfrac{\Gamma\left
(\frac{\alpha' k_{(2,1)}^\rho k_{{(2,1)}\rho} -3}{4}\right)
\Gamma\left(1-\alpha' \left[11p_\rho
p_\sigma S_{(1,2)}^{\rho\sigma} - 2 p_a p_a\right]
\right)} {\Gamma\left( 1+ \frac{\alpha'
k_{(2,1)}^\rho k_{{(2,1)}\rho}
-3}{4} -\alpha \left[11p_\rho p_\sigma
S_{(1,2)}^{\rho\sigma} - 2 p_a p_a\right]
\right)}\right.
\nonumber \\
&+& 11\alpha' p_\rho p_\sigma
S_{(1,2)}^{\rho\sigma} \left.\dfrac{\Gamma\left(\frac{\alpha'
k_{(2,1)}^\rho k_{{(2,1)}\rho} -3}{4}\right) \Gamma
\left(-\alpha' \left[11p_\rho
p_\sigma S_{(1,2)}^{\rho\sigma}
- 2 p_a p_a\right]\right)} {\Gamma\left(
\frac{\alpha' k_{(2,1)}^\rho k_{{(2,1)}\rho} -3}{4}
-\alpha' \left[11p_\rho p_\sigma S_{(1,2)}^{\rho\sigma}
- 2 p_a p_a\right]\right)}\right\}.
\label{4.7}
\end{eqnarray}
For the untwisted sector, we obtain
\begin{eqnarray}
\int_0^{\infty} dt' \int_0^\infty
d\tau e^{- t' \alpha'k_1^\mu k_{1 \mu} }
e^{- \tau \alpha'k_2^\mu k_{2 \mu} }
e^{4(t'+\tau)}\langle e^{ip\cdot X_{\rm osc}}
\rangle^{\rm U}|_{t\rightarrow\infty}^{\hat{\mathcal{S}}
= \mathbf{1}} = \mathcal{J}_1 \mathcal{J}_2
\label{4.8}
\end{eqnarray}
\begin{eqnarray}
\mathcal{J}_{1,2} &=& \dfrac{1}{4}\exp
\left[13 \alpha' p_\mu p_\nu
\hat S_{(1,2)}^{\mu\nu}(\gamma_{\rm EM}-1)\right]
\left\{ \dfrac{\Gamma\left
(\frac{\alpha' k_{(2,1)}^\mu k_{{(2,1)}\mu} -4}{4}\right)
\Gamma\left(1-13\alpha' p_\mu
p_\nu \hat S_{(1,2)}^{\mu\nu}
\right)} {\Gamma\left( 1+ \frac{\alpha'
k_{(2,1)}^\mu k_{{(2,1)}\mu}
-4}{4} -13\alpha' p_\mu p_\nu
\hat S_{(1,2)}^{\mu\nu}\right)}\right.
\nonumber \\
&+& 13 \alpha'  p_\mu p_\nu
\hat S_{(1,2)}^{\mu\nu} \left.
\dfrac{\Gamma\left(\frac{\alpha'
k_{(2,1)}^\mu k_{{(2,1)}\mu} -4}{4}\right) \Gamma
\left(-13\alpha' p_\mu p_\nu \hat S_{(1,2)}^{\mu\nu}
\right)} {\Gamma\left(
\frac{\alpha' k_{(2,1)}^\mu k_{{(2,1)}\mu} -4}{4}
-13\alpha' p_\mu p_\nu \hat S_{(1,2)}^{\mu\nu}
\right)}\right\}.
\label{4.9}
\end{eqnarray}
These are the typical factors, which emerge
in the two-point functions
on the worldsheets with the disk topology
(see e.g. \cite{54}-\cite{57}).

\subsection{The graviton emission}
\label{401}

For obtaining a generalized formulation,
we do not impose extra assumptions to the
polarization tensor.
The polarization tensors should
satisfy the condition $p^\mu \epsilon_{\mu\nu} = 0$.
The polarization tensor of the graviton
is symmetric and traceless, i.e.,
$\epsilon^{\rm g}_{\mu\nu} = \epsilon^{\rm g}_{\nu\mu}$
and $\epsilon^{{\rm g}\ \mu}_\mu = 0$.
Therefore, the untwisted part of the
radiation amplitude of the graviton is
given by Eq. \eqref{4.1} with the following variables
\begin{eqnarray}
\mathbf{A}^{\rm U} &=& \epsilon^{\rm g}_{\alpha\beta}
\left[ p_\xi p_\theta \left(Q_1^{\alpha\beta}
Q_1^{\xi \theta} - Q_1^{T \alpha\xi} Q_1^{T \beta\theta}
\right)+ 2 p_\xi p^\beta Q_1^{T \alpha\xi}
+ p^Ip_I Q_1^{\alpha\beta}\right]
\nonumber \\
&+& \epsilon^{\rm g}_{IJ} \left[ \delta^{IJ}
\left(p_\xi p_\theta Q_1^{\xi \theta}
+ p^Kp_K \right) - p^I p^J\right],
\label{4.10}
\end{eqnarray}
\begin{eqnarray}
\mathbf{B}^{\rm U} &=& \epsilon^{\rm g}_{\alpha\beta}
\left[ p_\xi p_\theta \left(Q_2^{ \alpha\beta}
Q_2^{ \xi \theta}- Q_2^{ \alpha\xi} Q_2^{ \beta\theta}
\right)+ 2 p_\xi p^\beta Q_2^{ \alpha\xi}
+ p^Ip_I Q_2^{ \alpha\beta}\right]
\nonumber \\
&+& \epsilon^{\rm g}_{IJ} \left[ \delta^{IJ}
\left(p_\xi p_\theta Q_2^{ \xi \theta}
+ p^Kp_K \right) - p^I p^J\right],
\label{4.11}
\end{eqnarray}
\begin{eqnarray}
\mathbf{C}^{\rm U} &=&\epsilon^{\rm g}_{\alpha\beta}
\left[ 26 \left(p^\alpha
+ \frac{1}{13}k_1^\alpha\right)\left(p^\beta
+ p_\theta Q_1^{T \beta\theta}\right)
- \frac{1}{2} Q_1^{\alpha\beta} (k_1^\mu k_{1\mu}
- k_2^\mu k_{2\mu})\right]
\nonumber \\
&-& \frac{1}{2} \epsilon_J^{{\rm g} J} \left[ 50 p^I p_I
+ 24 p_\alpha p_\beta Q_1^{\alpha\beta}
+ 26 p^\alpha p_\alpha + k_1^\mu k_{1\mu}
- k_2^\mu k_{2\mu} \right]
\nonumber \\
&+& \epsilon_{IJ}^{\rm g} (2 k_1^I p^J
- 26 p^I p^J) ,
\label{4.12}
\end{eqnarray}
\begin{eqnarray}
\mathbf{D}^{\rm U} &=&\epsilon^{\rm g}_{\alpha\beta}
\left[ 26 \left(p^\alpha + \frac{1}{13}k_1^\alpha\right)
\left(p^\beta+ p_\theta Q_2^{ \beta\theta}\right)
- \frac{1}{2} Q_2^{\alpha\beta} (k_1^\mu k_{1\mu}
- k_2^\mu k_{2\mu})\right]
\nonumber \\
&-& \frac{1}{2} \epsilon_J^{{\rm g} J} \left[ 50 p^I p_I
+ 24 p_\alpha p_\beta Q_2^{\alpha\beta} + 26 p^\alpha p_\alpha
+ k_1^\mu k_{1\mu} - k_2^\mu k_{2\mu} \right]
\nonumber \\
&+& \epsilon_{IJ}^{\rm g} (2 k_1^I p^J - 26 p^I p^J),
\label{4.13}
\end{eqnarray}
\begin{eqnarray}
\mathbf{E}^{\rm U} &=& \epsilon_{\alpha\beta}^{\rm g}
\left[ \frac{13}{2} \eta^{\alpha\beta} (k_1^\mu k_{1\mu}
- k_2^\mu k_{2\mu}) + (26p - k_1)^\alpha k_1^\beta
- 169 p^\alpha p^\beta\right]
\nonumber \\
&+& 13 \ \epsilon_{IJ}^{\rm g} \left[ \left( k_1
- 13 p\right)^I \left( p - \frac{1}{13}k_1\right)^J
+ \frac{1}{2} \delta^{IJ}  \right]
\nonumber \\
&+& 26 \ \epsilon^{\rm g}_{\alpha I}
\left( p- \frac{1}{13}k_1\right)^\alpha
\left( k_1+ 13 p\right)^I .
\label{4.14}
\end{eqnarray}
For the twisted sector, the graviton emission
amplitude is described by Eq. \eqref{4.2}
with the following variables
\begin{eqnarray}
\mathbf{A}^{\rm T} &=& \epsilon^{\rm g}_{\alpha\beta}
\left[ p_\xi p_\theta \left(Q_1^{\alpha\beta}
Q_1^{\xi \theta} - Q_1^{T \alpha\xi} Q_1^{T \beta\theta}
\right)+ 2 p_\xi p^\beta Q_1^{T \alpha\xi}
+ p^i p_i Q_1^{\alpha\beta}\right]
\nonumber \\
&+& \epsilon^{\rm g}_{ij} \left[ \delta^{ij}
\left(p_\alpha p_\beta Q_1^{\alpha\beta}
+ p^kp_k \right) - p^i p^j\right]
+ 2 \epsilon_{a \alpha}^{\rm g}
\ p^a p_\beta Q_1^{T \alpha\beta},
\label{4.15}
\end{eqnarray}
\begin{eqnarray}
\mathbf{B}^{\rm T} &=& \epsilon^{\rm g}_{\alpha\beta}
\left[ p_\xi p_\theta \left(Q_2^{ \alpha\beta}
Q_2^{ \xi \theta} - Q_2^{ \alpha\xi} Q_2^{ \beta\theta}
\right)+ 2 p_\xi p^\beta Q_2^{ \alpha\xi}
+ p^i p_i Q_2^{ \alpha\beta}\right]
\nonumber \\
&+& \epsilon^{\rm g}_{ij} \left[ \delta^{ij}
\left(p_\alpha p_\beta Q_2^{\alpha\beta}
+ p^kp_k \right) - p^i p^j\right] + 2
\epsilon_{a \alpha}^{\rm g} \ p^a p_\beta
Q_2^{ \alpha\beta},
\label{4.16}
\end{eqnarray}
\begin{eqnarray}
\mathbf{C}^{\rm T} &=& 22
\epsilon_{\alpha\beta}^{\rm g}
\left[ \left(p^\alpha + \frac{1}{11}
k_1^\alpha\right)\left(p^\beta
+ p_\theta Q_1^{T \beta\theta}\right)
- \frac{1}{44}Q_1^{\alpha\beta} (k_1^\rho k_{1\rho}
- k_2^\rho k_{2\rho})\right]
\nonumber \\
&+& \epsilon^{{\rm g} \ a}_a
\left(p_\alpha p_\beta Q_1^{\alpha\beta}
+ p^i p_i\right) - 22\epsilon^{\rm g}_{a \alpha}
p^a \left( p+ \frac{1}{11} k_1\right)^\alpha
+ \epsilon_{ij}^{\rm g} \left(2 k_1^i p^j
- 22 p^ip^j\right)
\nonumber \\
&-& \epsilon^{{\rm g} \ i}_i \left[ 21
p^j p_j + k_1^\rho k_{1\rho}
- k_2^\rho k_{2\rho} + 11 p_\alpha p_\beta
\left( Q_1^{\alpha\beta} + \eta^{\alpha\beta}\right)
- p_\alpha p_\beta  Q_1^{\alpha\beta}\right],
\label{4.17}
\end{eqnarray}
\begin{eqnarray}
\mathbf{D}^{\rm T} &=& 22
\epsilon_{\alpha\beta}^{\rm g}
\left[ \left(p^\alpha + \frac{1}{11}
k_1^\alpha\right)\left(p^\beta
+ p_\theta Q_2^{ \beta\theta}\right)
- \frac{1}{44}Q_2^{\alpha\beta}
(k_1^\rho k_{1\rho}
- k_2^\rho k_{2\rho})\right]
\nonumber \\
&+& \epsilon^{{\rm g} \ a}_a \left(p_\alpha
p_\beta Q_2^{\alpha\beta}
+ p^i p_i\right) - 22
\epsilon^{\rm g}_{a \alpha} p^a \left( p
+ \frac{1}{11} k_1\right)^\alpha
+ \epsilon_{ij}^{\rm g} \left(2 k_1^i p^j
- 22 p^ip^j\right)
\nonumber \\
&-& \epsilon^{{\rm g} \ i}_i \left[ 21 p^j p_j
+ k_1^\rho k_{1\rho} - k_2^\rho k_{2\rho}
+ 11 p_\alpha p_\beta \left( Q_2^{\alpha\beta}
+ \eta^{\alpha\beta}\right) - p_\alpha p_\beta
Q_2^{\alpha\beta}\right],
\label{4.18}
\end{eqnarray}
\begin{eqnarray}
\mathbf{E}^{\rm T} &=& \epsilon_{\alpha\beta}^{\rm g}
\left[ \frac{11}{2} \eta^{\alpha\beta} (k_1^\rho k_{1\rho}
- k_2^\rho k_{2\rho}) + \left(22p
- k_1\right)^\alpha k_1^\beta
- \frac{121}{2} p^\alpha p^\beta\right]
\nonumber \\
&+& 11 \ \epsilon_{ij}^{\rm g} \left[ \left( k_1
- 11 p\right)^i
\left( p - \frac{1}{11}k_1\right)^j
+ \frac{1}{2} \delta^{ij}  \right]
\nonumber \\
&+& 11\ \epsilon^{\rm g}_{\alpha i}
\left( p- \frac{1}{11}k_1\right)^\alpha
\left( k_1+ 11 p\right)^i,
\label{4.19}
\end{eqnarray}
\begin{eqnarray}
\mathbf{F} &=& \epsilon^{\rm g}_{aa'}
\left(\delta^{aa'} p^b p_b
- p^a p^{a'}\right) , \qquad \mathbf{G}
=-\frac{1}{2} \epsilon_{a}^{{\rm g} \ a'}(k_1^\rho k_{1\rho}
- k_2^\rho k_{2\rho}).
\label{4.20}
\end{eqnarray}

\subsection{The Kalb-Ramond emission}
\label{402}

The Kalb-Ramond (axion) polarization tensor
is anti-symmetric $\epsilon^x_{\mu\nu} =
- \epsilon^x_{\nu\mu}$. According to the
right-hand sides of
Eqs. \eqref{3.35}, \eqref{3.37} and \eqref{3.38},
which are symmetric, the amplitude of the axion emission
should be independent of the matrix elements
$\epsilon^x_{IJ}$, $\epsilon^x_{ij}$ and
$\epsilon^x_{ab}$. Hence, for the untwisted sector,
we acquire
\begin{eqnarray}
\mathbf{A}^{\rm U} &=& -\epsilon^x_{\alpha\beta}
\left[ p_\xi p_\theta \left(Q_1^{\alpha\beta}
Q_1^{\xi \theta} + Q_1^{T\alpha\xi} Q_1^{T\beta\theta}
\right)+ p^Ip_I Q_1^{\alpha\beta}\right] ,
\label{4.21}\\
\mathbf{B}^{\rm U} &=& \epsilon^x_{\alpha\beta}
\left[ p_\xi p_\theta \left(Q_2^{ \alpha\beta}
Q_2^{ \xi \theta}- Q_2^{ \alpha\xi} Q_2^{ \beta\theta}
\right)+ p^Ip_I Q_2^{ \alpha\beta}\right],
\label{4.22}\\
\mathbf{C}^{\rm U} &=&\mathbf{D}^{\rm U} =
\mathbf{E}^{\rm U} = 0 .
\label{4.23}
\end{eqnarray}
For the twisted sector, we have
\begin{eqnarray}
\mathbf{A}^{\rm T} &=& -\epsilon^x_{\alpha\beta}
\left[ p_\xi p_\theta \left(Q_1^{\alpha\beta}
Q_1^{\xi \theta} + Q_1^{T\alpha\xi} Q_1^{T\beta\theta}
\right)+ p^ip_i Q_1^{\alpha\beta}\right] ,
\label{4.24}\\
\mathbf{B}^{\rm T} &=& \epsilon^x_{\alpha\beta}
\left[ p_\xi p_\theta \left(Q_2^{ \alpha\beta}
 Q_2^{ \xi \theta}- Q_2^{ \alpha\xi} Q_2^{ \beta\theta}
\right)+ p^ip_i Q_2^{ \alpha\beta}\right],
\label{4.25} \\
\mathbf{C}^{\rm T} &=&\mathbf{D}^{\rm T} =
\mathbf{E}^{\rm T} = \mathbf{F} = \mathbf{G} \ = 0.
\label{4.26}
\end{eqnarray}

\subsection{The dilaton emission}
\label{403}

In contrast to the graviton's and
axion's polarization tensors,
the explicit form of the dilation
polarization tensor is available
\begin{equation}
\label{4.27}
\epsilon^\phi_{\mu\nu} = \frac{1}{\sqrt{24}}
\left( \eta_{\mu\nu} - p_\mu \bar
p_\nu - p_\nu \bar p_\mu\right),
\end{equation}
where ${\bar p}^\mu$ is a light-like vector,
i.e. $\bar p^\mu \bar p_\mu =0$,
such that $p^\mu \bar p_\mu =1$.
Thus, the untwisted sector of the dilaton emission
possesses the following variables
\begin{eqnarray}
\mathbf{A}^{\rm U} &=& \frac{1}{\sqrt{24}}
\left\{ p_\xi p_\theta
\left(Q_{1 \ \alpha}^{\alpha }Q_1^{\xi \theta}
- Q_1^{T \alpha\xi } Q_{1 \alpha}^{T \ \theta}\right)
+ 2 p_\alpha p_\beta Q_1^{\alpha\beta}
+ p^I p_I Q_{1\ \alpha}^{\alpha}
\nonumber
\right.\\
&+& (p_\alpha \bar p_\beta - p_\beta \bar p_\alpha)
\left[ p_\xi p_\theta
\left(Q_1^{\alpha\beta }Q_1^{\xi \theta}
+ Q_1^{T \alpha\xi } Q_{1 }^{T \beta\theta}\right)
- 2 p_\xi p^\beta Q_1^{T \alpha\xi}
+ p^I p_I Q_{1}^{\alpha\beta}\right]
\nonumber \\
&+&\left. \left[(25-p)\left( p_\alpha p_\beta Q_1^{\alpha\beta}
+ p^I p_I\right) - p^I p_I\right] - 2 p_I \bar p^I
p_\alpha p_\beta Q_1^{\alpha\beta}\right\},
\label{4.28}
\end{eqnarray}
\begin{eqnarray}
\mathbf{B}^{\rm U} &=& \frac{1}{\sqrt{24}}
\left\{ p_\xi p_\theta
\left(Q_{2 \ \alpha}^{ \alpha }Q_2^{ \xi \theta}
- Q_2^{ \alpha\xi } Q_{2 \alpha}^{ \ \theta}\right)
+ 2 p_\alpha p_\beta Q_2^{\alpha\beta}
+ p^I p_I Q_{2\ \alpha}^{ \alpha}
\nonumber
\right.\\
&-& (p_\alpha \bar p_\beta - p_\beta \bar p_\alpha)
\left[ p_\xi p_\theta
\left(Q_2^{ \alpha\beta }Q_2^{ \xi \theta}
- Q_2^{ \alpha\xi } Q_{2 }^{ \beta\theta}\right)
+ 2 p_\xi p^\beta Q_2^{ \alpha\xi}
+ p^I p_I Q_{2}^{ \alpha\beta}\right]
\nonumber \\
&+&\left. \left[(25-p)
\left( p_\alpha p_\beta Q_2^{\alpha\beta}
+ p^I p_I\right) - p^I p_I\right]
- 2 p_I \bar p^I p_\alpha p_\beta
Q_2^{\alpha\beta}\right\},
\label{4.29}
\end{eqnarray}
\begin{eqnarray}
\mathbf{C}^{\rm U} &=& \frac{1}{\sqrt{24}}
\left\{ 26\left( p_\beta
Q_{1 \  \alpha}^{T \beta} + p_\alpha\right)
\left( p + \frac{1}{13} k_1\right)^\alpha - \frac{1}{2}
Q_{1 \ \alpha}^{\alpha} (k_1^\mu k_{1\mu}
- k_2^\mu k_{2\mu}) \right.
\nonumber \\
&-& (p_\alpha \bar p_\beta - p_\beta \bar p_\alpha)
\left [ 26\left( p_\xi Q_1^{T \beta\xi}
+ p^\beta\right)\left(p
+ \frac{1}{13}  k_1\right)^\alpha + \frac{1}{2}
Q_1^{\alpha\beta} (k_1^\mu k_{1\mu}
- k_2^\mu k_{2\mu})\right]
\nonumber \\
&+& 2 p_I\left(k_1^I -13 p^I\right)
\left(1- 2 \bar p_J p^J\right)
- (p+1) \left(p_\alpha p_\beta Q_1^{\alpha\beta}
+ p^I p_I\right)
\nonumber \\
&+& 2 p_\alpha \bar p^\alpha \left(p_\xi p_\theta
Q_1^{\xi \theta} + p^I p_I\right) + (25-p- 2 p_I \bar p^I)
\left[ 26 p^J p_J +\frac{1}{2} (k_1^\mu k_{1\mu}
- k_2^\mu k_{2\mu})\right.
\nonumber \\
&+& \left.\left. 13
p_\alpha p_\beta \left(Q_1^{\alpha\beta}
+ \eta^{\alpha\beta}\right)\right]\right\},
\label{4.30}
\end{eqnarray}
\begin{eqnarray}
\mathbf{D}^{\rm U} &=& \frac{1}{\sqrt{24}}
\left\{ 26\left( p_\beta Q_{2 \  \alpha}^{\beta}
+ p_\alpha\right) \left( p + \frac{1}{13} k_1\right)^\alpha
- \frac{1}{2} Q_{2 \ \alpha}^{ \alpha} (k_1^\mu k_{1\mu}
- k_2^\mu k_{2\mu}) \right.
\nonumber \\
&-& (p_\alpha \bar p_\beta - p_\beta \bar p_\alpha)
\left [ 26\left( p_\xi Q_2^{ \beta\xi}
+ p^\beta\right)\left(p
+ \frac{1}{13}  k_1\right)^\alpha
-\frac{1}{2} Q_2^{\alpha\beta}
(k_1^\mu k_{1\mu} - k_2^\mu k_{2\mu})\right]
\nonumber \\
&+& 2 p_I\left(k_1^I -13 p^I\right)
\left(1- 2 \bar p_J P^J\right) - (p+1)
\left(p_\alpha p_\beta Q_2^{\alpha\beta}
+ p^I p_I\right)
\nonumber \\
&+& 2 p_\alpha \bar p^\alpha
\left(p_\xi p_\theta Q_2^{ \xi \theta}
+ p^I p_I\right) + (25-p
- 2 p_I \bar p^I) \left[ 26 p^J p_J
+\frac{1}{2} (k_1^\mu k_{1\mu} - k_2^\mu k_{2\mu})\right.
\nonumber \\
&+& \left.\left.13 p_\alpha p_\beta
\left( Q_2^{\alpha\beta} + \eta^{\alpha\beta}\right)\right]
\right\},
\label{4.31}
\end{eqnarray}
\begin{eqnarray}
\mathbf{E}^{\rm U} &=& \frac{1}{\sqrt{24}}\left\{
\frac{13(p+1)}{2} (k_1^\mu k_{1\mu} - k_2^\mu k_{2\mu})
+ \left(26 p +k_1\right)^\alpha k_{1 \alpha}
+ 169 p^\alpha p_\alpha \right.
\nonumber \\
&-&  2 p_\alpha \bar p_\beta \left( \frac{13}{2}
\eta^{\alpha\beta} (k_1^\mu k_{1\mu} - k_2^\mu k_{2\mu})
+ (26 p + k_1)^\alpha k_1^\beta + 169
p^\alpha p^\beta \right)
\nonumber\\
&+& 13 \left[ \frac{25-p}{2}
(k_1^\mu k_{1\mu} - k_2^\mu k_{2\mu})
 +  \left(p - \frac{1}{13} k_1\right)_I
 \left(k_1 - 13 p\right)^I \right]
\nonumber \\
 &+&\left. \left[ 26 p_\alpha - 52
p_\alpha \bar p_\beta p^\beta
 +k_1^I (p_\alpha \bar p_I + p_I \bar p_\alpha)
\right] \left( k_1 + 13 p\right)^\alpha \right\}.
\label{4.32}
\end{eqnarray}
For the twisted sector we have
\begin{eqnarray}
\mathbf{A}^{\rm T} &=& \frac{1}{\sqrt{24}}
\left\{ p_\xi p_\theta
\left(Q_{1 \ \alpha}^{\alpha }Q_1^{\xi \theta}
- Q_1^{T \alpha\xi } Q_{1 \alpha}^{T \ \theta}\right)
+ 2 p_\alpha p_\beta Q_1^{\alpha\beta}
+ p^i p_i Q_{1\ \alpha}^{\alpha}
\nonumber
\right.\\
&+& (p_\alpha \bar p_\beta - p_\beta \bar p_\alpha)
\left[ p_\xi p_\theta
\left(Q_1^{T \alpha\beta }Q_1^{T \xi \theta}
+ Q_1^{T \alpha\xi } Q_{1 }^{T \beta\theta}\right)
- 2 p_\xi p^\beta Q_1^{T \alpha\xi}
+ p^i p_i Q_{1}^{\alpha\beta}\right]
\nonumber \\
&+& \left[(21-p)\left( p_\alpha p_\beta Q_1^{\alpha\beta}
+ p^i p_i\right) - p^i p_i\right]
\nonumber \\
&-& \left. 2 p_\xi \left[ p^a
\left(p_a \bar p_\alpha + p_\alpha \bar p_a \right)
+ p^i \bar p_i p_\alpha\right]
Q_1^{T\alpha\xi}  \right\},
\label{4.33}
\end{eqnarray}
\begin{eqnarray}
\mathbf{B}^{\rm T} &=& \frac{1}{\sqrt{24}}
\left\{ p_\xi p_\theta
\left(Q_{2 \ \alpha}^{ \alpha }Q_2^{\xi \theta}
- Q_2^{ \alpha\xi } Q_{2 \alpha}^{ \ \theta}\right)
+ 2 p_\alpha p_\beta Q_2^{\alpha\beta}
+ p^i p_i Q_{2\ \alpha}^{ \alpha}
\nonumber
\right.\\
&-& (p_\alpha \bar p_\beta - p_\beta \bar p_\alpha)
\left[ p_\xi p_\theta
\left(Q_2^{ \alpha\beta }Q_2^{\xi\theta}
- Q_2^{ \alpha\xi } Q_{2 }^{ \beta\theta}\right)
+ 2 p_\xi p^\beta Q_2^{ \alpha\xi}
+ p^i p_i Q_{2}^{ \alpha\beta}\right]
\nonumber \\
&+& \left[(21-p)\left( p_\alpha p_\beta Q_2^{\alpha\beta}
+ p^i p_i\right) - p^i p_i\right]
\nonumber \\
&-& \left. 2 p_\xi \left[ p^a
\left(p_a \bar p_\alpha + p_\alpha \bar p_a \right)
+ p^i \bar p_i p_\alpha\right] Q_2^{\alpha\xi}  \right\},
\label{4.34}
\end{eqnarray}
\begin{eqnarray}
\mathbf{C}^{\rm T} &=& \frac{1}{\sqrt{24}}
\left\{ 22 \left( p_\beta
Q_{1 \  \alpha}^{T\beta} + p_\alpha\right)
\left( p + \frac{1}{11} k_1\right)^\alpha - \frac{1}{2}
Q_{1 \ \alpha}^{\alpha} (k_1^\rho k_{1\rho}
- k_2^\rho k_{2\rho}) \right.
\nonumber \\
&-& (p_\alpha \bar p_\beta - p_\beta \bar p_\alpha)
\left [ 22 \left( p_\xi Q_1^{T \beta\xi}
+ p^\beta\right)\left(p
+ \frac{1}{11}  k_1\right)^\alpha
\right.
\nonumber \\
&+&\left. \frac{1}{2}Q_1^{\alpha\beta} (k_1^\rho k_{1\rho}
- k_2^\rho k_{2\rho})\right]+ 2p_i( k_1^i - 11 p^i)
(1-2 {\bar p_j}p^j )
\nonumber \\
&+& (2p_\alpha {\bar p}^\alpha- p-1) \left(p_\xi
p_\beta Q_1^{\xi\beta} + p^i p_i\right)
\nonumber \\
&+& (21-p - 2p_i \bar p^i)
\left[ 22 p^j p_j +\frac{1}{2} (k_1^\rho k_{1\rho}
- k_2^\rho k_{2\rho})\right.
\nonumber \\
&+& \left. 11 p_\alpha p_\beta
\left( Q_1^{\alpha\beta} +
\eta^{\alpha\beta}\right)\right]
- \left. 22 p^a (p_\alpha \bar
p_a + p_a \bar p_\alpha)
\left( p + \frac{1}{11} k_1\right)^\alpha \right\},
\label{4.35}
\end{eqnarray}
\begin{eqnarray}
\mathbf{D}^{\rm T} &=& \frac{1}{\sqrt{24}}
\left\{ 22 \left( p_\xi Q_{2 \  \alpha}^{\xi}
+ p_\alpha\right)
\left( p + \frac{1}{11} k_1\right)^\alpha - \frac{1}{2}
Q_{2 \ \alpha}^{ \alpha} (k_1^\rho k_{1\rho}
- k_2^\rho k_{2\rho}) \right.
\nonumber \\
&-& (p_\alpha \bar p_\beta - p_\beta \bar p_\alpha)
\left [ 22 \left( p_\xi
Q_2^{ \beta\xi} + p^\beta\right)\left(p
+ \frac{1}{11}  k_1\right)^\alpha
\right.
\nonumber \\
&-&\left. \frac{1}{2}
Q_2^{ \alpha\beta} (k_1^\rho k_{1\rho}
- k_2^\rho k_{2\rho})\right]
+ 2p_i( k_1^i - 11 p^i)(1-2 {\bar p_j}p^j )
\nonumber \\
&+& (2 p_\alpha {\bar p}^\alpha - p-1) \left(p_\xi
p_\beta Q_2^{\xi\beta} + p^i p_i\right)
\nonumber \\
&+& (21-p - 2 p_i \bar p^i)
\left[ 22 p^j p_j +\frac{1}{2} (k_1^\rho k_{1\rho}
- k_2^\rho k_{2\rho})\right.
\nonumber \\
&+& \left. 11 p_\alpha p_\beta
\left( Q_2^{\alpha\beta} + \eta^{\alpha\beta}\right)
\right]- \left. 22 p^a (p_\alpha
\bar p_a + p_a \bar p_\alpha)
\left( p + \frac{1}{11}
k_1\right)^\alpha \right\},
\label{4.36}
\end{eqnarray}
\begin{eqnarray}
\mathbf{E}^{\rm T} &=& \frac{1}{\sqrt{24}}\left\{
\frac{11(p+1)}{2} (k_1^\rho k_{1\rho} - k_2^\rho k_{2\rho})
+ \left(22 p +k_1\right)^\alpha k_{1 \alpha}
+ \frac{121}{2}p^\alpha p_\alpha \right.
\nonumber \\
&-&  2 p_\alpha \bar p_\beta \left( \frac{11}{2}
\eta^{\alpha\beta} (k_1^\rho k_{1\rho} - k_2^\rho k_{2\rho})
+ \left(22 p + k_1\right)^\alpha k_1^\beta
+ \frac{121}{2} p^\alpha p^\beta \right)
\nonumber\\
&+& 11 \left[ \frac{21-p}{2}
(k_1^\rho k_{1\rho} - k_2^\rho k_{2\rho})
+  \left(p - \frac{1}{11} k_1\right)_i
\left(k_1 - 11 p\right)^i \right]
\nonumber \\
&+&\left. \left[ 22 p_\alpha
- 44 p_\alpha \bar p_\beta p^\beta
+k_1^i \left(p_\alpha \bar p_i + p_i \bar p_\alpha\right)
\right]\left( k_1 +11 p\right)^\alpha \right\},
\label{4.37}
\end{eqnarray}
\begin{eqnarray}
\mathbf{F} &=& \frac{1}{\sqrt{24}}p^a p_a
\left( 3 -4 p_b\bar p^b\right),
\label{4.38}
\end{eqnarray}
\begin{eqnarray}
\mathbf{G} &=& -\frac{1}{\sqrt{24}} \left(2
- p_a \bar p^a\right) (k_1^\rho k_{1\rho}
- k_2^\rho k_{2\rho}).
\label{4.39}
\end{eqnarray}

Substituting the variables
$\mathbf{A}^{\rm U/T}$,
$\mathbf{B}^{\rm U/T}$,
$\mathbf{C}^{\rm U/T}$,
$\mathbf{D}^{\rm U/T}$,
$\mathbf E^{\rm U/T}$,
$\mathbf F$ and $\mathbf G$, from
Eqs. \eqref{4.10}-\eqref{4.26} and
Eqs. \eqref{4.28}-\eqref{4.39},
into the partial amplitudes \eqref{4.1}
and \eqref{4.2} yields the total emission amplitudes
of the graviton, Kalb-Ramond and dilaton
states.

\subsection{Physical interpretation}
\label{404}

In order to extract the physical
interpretations from these amplitudes,
let us evaluate our configuration at low energy limit (LEL).
Therefore, the solutions of the proper
times integrals in both sectors, i.e.
Eqs. \eqref{4.7} and \eqref{4.9}, reduce to
\begin{eqnarray}
\mathcal{I}_1 &\approx& \frac{1}{\alpha'
k_2^\rho k_{2\rho} -3} ,
\nonumber \\
\mathcal{I}_2 &\approx& \frac{1}{\alpha'
k_1^\rho k_{1\rho} -3} ,
\nonumber \\
\mathcal{J}_1 &\approx& \frac{1}{\alpha'
k_2^\mu k_{2\mu} -4} ,
\nonumber \\
\mathcal{J}_2 &\approx& \frac{1}{\alpha'
k_1^\mu k_{1\mu} -4}.
\label{4.40}
\end{eqnarray}
Now let us apply the constant shifts
$3/{\alpha'} $ and $4/{\alpha'}$
to the squared momenta in the twisted and untwisted
sectors, respectively.
Adding all these together gives the following
amplitudes for the low energy limit case
\begin{eqnarray}
\mathcal{A}^{\rm U}_{\rm LEL}
|_{t \rightarrow \infty }^{\hat{\mathcal{S}}
= \mathbf{1}} = &=& \dfrac{T_p^2\sqrt{\det(M_1 M_2)}}{8
(2\pi)^{23-2p}} \frac{1}{|v_1- v_2|}
\prod_{\bar\alpha =1}^p \delta
(p^{\bar\alpha})
\nonumber \\
&\times& \int_{-\infty}^{+\infty}
\prod_{I \ne i_v}^{25} dk_1^I e^{i k_1^I b_I}
\left\{ \frac{\mathbf{E}^{\rm U}_{\rm shift}}
{k_1^\mu k_{1\mu} k^\nu_2 k_{2\nu}}
-\frac{\mathcal{X}^{\rm U}_{\rm shift}}{k_1^\mu k_{1\mu}}
-\frac{\mathcal{Y}^{\rm U}_{\rm shift}}{k_2^\mu k_{2\mu}}
\right\},
\label{4.41}
\end{eqnarray}
for the untwisted sector, with
\begin{eqnarray}
\mathcal{X}^{\rm U}_{\rm shift}
&=& \frac{1}{2 p_\mu p_\nu \hat S_1^{\mu\nu}}
\left(\mathbf{C}_{\rm shift}^{\rm U}
- \frac{k^\mu_2 k_{2\mu}}{2 p_\mu p_\nu \hat S_1^{\mu\nu}}
\mathbf{A}_{\rm shift}^{\rm U}\right),
\nonumber \\
\mathcal{Y}^{\rm U}_{\rm shift}
&=& \frac{1}{2 p_\mu p_\nu \hat S_2^{ \mu\nu}}
\left(\mathbf{D}_{\rm shift}^{\rm U}
- \frac{k^\mu_1 k_{1\mu}}{2 p_\mu p_\nu \hat S_2^{ \mu\nu}}
\mathbf{B}_{\rm shift}^{\rm U}\right).
\label{4.42}
\end{eqnarray}
For the twisted sector, we find
\begin{eqnarray}
\mathcal{A}^{\rm T}_{\rm LEL}
|_{t \rightarrow \infty }^{\mathcal{S}
= \mathbf{1}} = &=& \dfrac{T_p^2\sqrt{\det(M_1 M_2)}}{8
(2\pi)^{19-2p}} \frac{1}{|v_1- v_2|}
\prod_{\bar\alpha =1}^p \delta
(p^{\bar\alpha})
\nonumber \\
&\times& \int_{-\infty}^{+\infty}
\prod_{i \ne i_v}^{21} dk_1^i e^{i k_1^i b_i}
\left\{ \frac{\mathbf{E}^{\rm T}_{\rm shift}}
{k_1^\rho k_{1\rho} k^\sigma k_{2\sigma}}
- \frac{\mathcal{X}^{\rm T}_{\rm shift}}
{k_1^\rho k_{1\rho}}
- \frac{\mathcal{Y}^{\rm T}_{\rm shift}}
{k_2^\rho k_{2\rho}}
\right\},
\label{4.43}
\end{eqnarray}
where
\begin{eqnarray}
\mathcal{X}^{\rm T}_{\rm shift}
&=& \frac{1}{2 p_\rho p_\sigma  S_1^{\rho\sigma}}
\left(\mathbf{C}_{\rm shift}^{\rm T}
- \frac{\mathbf{A}_{\rm shift}^{\rm T}
k^\rho_2 k_{2\rho}}{2 p_\rho p_\sigma
S_1^{\rho\sigma}} \right)
+ \frac{1}{2 p_a p_a}\left(\mathbf{G}_{\rm shift}
- \frac{\mathbf{F}_{\rm shift} k^\rho_2 k_{2\rho}}
{2 p_a p_a} \right) ,
\nonumber \\
\mathcal{Y}^{\rm T}_{\rm shift}
&=& \frac{1}{2 p_\rho p_\sigma  S_2^{ \rho\sigma}}
\left(\mathbf{D}_{\rm shift}^{\rm T}
- \frac{\mathbf{B}_{\rm shift}^{\rm T}
k^\rho k_{1\rho}}{2 p_\rho p_\sigma  S_2^{ \rho\sigma}}
\right)+ \frac{1}{2 p_a p_a}
\left[\mathbf{G}_{\rm shift}
- \frac{\mathbf{F}_{\rm shift} k^\rho_1 k_{1\rho}}
{2 p_a p_a} \right).
\label{4.44}
\end{eqnarray}
The summation of the amplitudes \eqref{4.41} and \eqref{4.43}
represents the low energy emission of one of the graviton,
Kalb-Ramond and dilaton states.
Our physical results are comparable with Ref. \cite{46}.

The denominators $ k_1^\mu k_{1\mu}$
and $ k_2^\mu k_{2\mu}$
(and similarly  $ k_1^\rho k_{1\rho}$
and $ k_2^\rho k_{2\rho}$)
correspond to the particle's propagator.
Besides, the terms with
$\mathcal{X}^{\rm T/U}_{\rm shift}$ and
$\mathcal{Y}^{\rm T/U}_{\rm shift}$
originate from a single-pole process.
We observe that the quantities
$\mathcal{X}^{\rm U}_{\rm shift}$ and
$\mathcal{X}^{\rm T}_{\rm shift}$
($\mathcal{Y}^{\rm U}_{\rm shift}$ and
$\mathcal{Y}^{\rm T}_{\rm shift}$)
completely depend on the velocity and fields of
the first (the second) brane. Thus,
the $\mathcal{X}^{\rm T/U}_{\rm shift}$-term
($\mathcal{Y}^{\rm T/U}_{\rm shift}$-term)
elaborates that a massless closed
string is emitted by the second (the first)
brane, then is absorbed by the first
(the second) brane, then after traveling
as an excited state, it re-decays by emitting a
massless particle on the first (second) brane.

The $\mathbf{E}^{\rm U}_{\rm shift}$- and
$\mathbf{E}^{\rm T}_{\rm shift}$-terms
denote the double-pole mechanism.
We saw that the quantities
$\mathbf{E}^{\rm U}_{\rm shift}$ and
$\mathbf{E}^{\rm T}_{\rm shift}$
are independent of the velocities and fields of both
branes. Hence, these terms clarify that
the radiation occurs from
the exchanged closed string
between the branes. The latter closed
string causes the interaction of the branes.

Since all quantities in
Eqs. \eqref{4.10}-\eqref{4.20} and in
Eqs. \eqref{4.28}-\eqref{4.39} are nonzero,
we conclude that each of the graviton and dilaton
can be emitted in one of the following
physical processes: radiation from the middle
points between the branes and or
emissions from each of the branes.
However, the vanishing value of
$\mathbf{E}^{\rm U/T}_{\rm shift}$
for the Kalb-Ramond state
reveals that the axion radiation cannot
occur in the middle region between the branes.
In other words, the axion emission
can take place only on the surface of
one of the branes.

\section{Conclusions}
\label{500}

We computed the boundary states,
which correspond to a fractional D$p$-brane,
in the untwisted and twisted sectors of
the bosonic string theory.
The fractional brane has been located at the
fixed-points of the non-compact orbifold
$\mathbb{R}^4/\mathbb{Z}_2$, and has
been dressed by the Kalb-Ramond field
and a $U(1)$ gauge potential.
Besides, it has a transverse motion.
The orbifold part of the spacetime, the internal and
background fields and dynamics of the brane drastically
affected the boundary states.

In the closed string channel, the interaction
of the D-branes happens via
the exchange of a closed string.
We inserted the general form of an
integrated vertex operator,
associated with a massless closed string, into the
interaction amplitude of the branes.
This defines the amplitude of the
closed string radiation from
the branes with an arbitrary distance.
The presence of various parameters in the setup
prominently generalized the radiation amplitude.
By adjusting the parameters, the
value of this amplitude can be
adjusted to any suitable configurations.

We obtained the radiation amplitude
for the case that the distance of the
interacting branes is very large.
To compare our results with the
Ref. \cite{46}, the assumptions
$\mathcal{S}=\hat{\mathcal{S}}=\mathbf{1}$ were imposed,
which yield some relations between the parameters
of our setup.
Then, we explicitly calculated the amplitude of the
graviton, Kalb-Ramond and dilaton emissions.
We observed that only one of the following radiations
could potentially take place: one radiation between
the interacting branes and two radiations
from the branes. All of these three possible
processes can occur for the dilaton and
graviton radiations, while the axion
radiation cannot occur in the middle region
of the branes.


\end{document}